\documentclass[12pt]{iopart}
\usepackage{graphicx}

\usepackage{iopams}
\usepackage{color}

\newcommand{\red}[1]{{#1}}

\begin{document}

\title{Spatial effect on stochastic dynamics of bistable evolutionary games}
\author{Kohaku H. Z. So$^1$, Hisashi Ohtsuki$^2$ and Takeo Kato$^3$}
\address{${}^1$ Departmenf of Physics, Graduate School of Science, The University of Tokyo, 7-3-1 Hongo, Bunkyo-ward, Tokyo 113-0033, Japan}
\address{${}^2$ Department of Evolutionary Studies of Biosystems, School of Advanced Sciences, 
	The Graduate University for Advanced Studies, Hayama, Kanagawa 240-0193, Japan}
\address{${}^3$ Institute for Solid State Physics, University of Tokyo, 5-1-5 Kashiwa-no-ha, Kashiwa, Chiba 277-8581, Japan}
\eads{7480926558@mail.ecc.u-tokyo.ac.jp}


\begin{abstract}
		We consider the \red{lifetimes} of metastable states in bistable evolutionary games \red{(coordination games)}, 
		and examine how they are affected by spatial structure. 
		A semiclassical approximation based on a path integral method is applied to 
		stochastic evolutionary game dynamics with and without spatial \red{structure}, 
		and the lifetimes of the metastable states are evaluated. 
		It is shown that the population dependence of the lifetimes is qualitatively different in these two models. 
		\red{Our result indicates that spatial structure can accelerate the transitions between metastable states.}
\end{abstract}
\pacs{05.40.-a, 87.23.Kg}

\noindent{\it Keywords\/}: Game-theory (Theory), Metastable states, Large deviations in non-equilibrium systems
\maketitle


\section{Introduction}
\label{sec:Intro}

Evolutionary game theory provides a mathematical framework for analyzing conflicts of interests between individuals. 
The theory is based on the principle of evolution, i.e. 
strategies that perform well in the population increase their abundance but those perform poorly are wiped out. 
It has been applied in many disciplines such as social sciences 
\cite{Weibull, Gintis2009} and evolutionary biology 
\cite{JMSbook,NowakBook,HofbauerSigmund}. 
Recently, methods developed in statistical physics 
have been utilised in researches of this interdisciplinary field \cite{SzaboToke,SzaboVukovSzolnoki,HauertSzabo,SzaboFath,AssafMobilia} 

Evolutionary game theory has often been \red{formulated} by a deterministic process by assuming an infinitely large population \cite{HofbauerSigmund,Gintis2000}. 
For example, replicator dynamics describe the deterministic change of frequencies of strategies \cite{TaylorJonker}. 
Resulting dynamics are, in most cases, frequency-dependent, 
which means that a winning strategy depends on the current distribution of strategies in the population.  

More recently, stochastic models have been applied to analyze evolutionary game dynamics in a finite population \cite{NowakBook,NowakSasakiTaylorFudenberg}. 
Examples include evolutionary game-theoretic versions of the Wright-Fisher and Moran processes 
\cite{TaylorFudenbergSasakiNowak,ImhofNowak}, 
which are historically known as models in population genetics \cite{Ewens}. 
Analyses of stochastic models reveal the effect of various random factors, such as demographic stochasticity or recursive mutations, on game dynamics. 

One of the most important applications of stochastic game dynamics is the study of transitions among multiple ``metastable states'' 
that would correspond to fixed points (hence equilibria) in a deterministic description. 
A metastable state of a stochastic process is \red{a state} where the system stays for a sufficiently long time. 
By applying stochastic game dynamics \red{and examining which metastable state has the longest lifetime}, 
one can reveal which metastable state is most likely to be realised 
\cite{NowakBook,KMR,Young,AntalEtal2009,TarnitaEtal2009,Ohtsuki2010}. 

Another dimension of studies in evolutionary game theory is the incorporation of  spatial structure into the model 
to explore the interplay between game and space \cite{SzaboToke,SzaboFath,Ellison1993,OhtsukiEtal2006,NowakEtal2010}. 
There are numerous studies in this direction, which suggest that the speed of evolution as well as its outcome can be significantly influenced by the presence of spatial structure. 

In this paper, we study how spatial \red{structure} affects the transition between metastable states in 
a class of evolutionary games called coordination games \cite{KMR}. 
In coordination games, there are two strategies, A and B. 
The payoff of an A-individual is higher when matched with another A-individual than when matched with a B-individual. 
Similarly, the payoff of a B-individual is higher when matched with another B-individual than when matched with an A-individual. 
\red{Cooperative hunting is one biological example of such games, 
where two hunters hunt either prey A or B; 
hunting the same prey together greatly improves the success, so coordination is the better strategy. 
This} game has bistability, because both the all-A state \red{($=$ everyone adopts strategy A)} and 
the all-B state \red{(=everyone adopts strategy B)} are \red{stable} fixed points of the \red{deterministic} evolutionary game dynamics. 
When a small ``mutation'', i.e. a small chance that the individuals will randomly change their strategy, is introduced, 
the two states (A-dominant and B-dominant) are realized as metastable states and transitions between them may occur, \red{i.e. they have finite lifetimes}. 
When the system is large, such transitions are extremely rare. 

To evaluate the \red{lifetimes} of metastable states, 
we adopt a semiclassical (WKB) approximation \cite{Dykman,ElgartKamenev,AssafMeerson2006-PRE,AssafMeerson2006-PRL,EscuderoKamenev,AltlandSimons}, 
which is an analog of the WKB approximation used in quantum mechanics. 
This method is known to be suitable for evaluating the probability of rare events caused by large fluctuations \cite{AltlandSimons}, 
because it can appropriately take into account an exponentially small tail of probability distribution 
beyond the Gaussian fluctuation considered in the Fokker-Planck approximation or the van Kampen expansion \cite{vanKampen}.  
The semiclassical approximation has been used for evaluating the probabilities of rare events 
in the context of extinction phenomena in ecology \cite{KesslerShnerb,AssafKamenevMeerson,AssafMeerson2010,OvaskainenMeerson,MeersonSasorov}. 
For evolutionary games, although it has been applied to the analysis of fixation \cite{AssafMobilia}, 
transition between metastable states, especially how it is affected by spatial structure, 
has not been studied well so far. 

Here we consider spatial effect on the transition between metastable states in coordination games. 
To investigate the spatial effect, we analyze two types of evolutionary game models: 
(i) a model without any spatial \red{structure} (the ``well-mixed'' model), and 
(ii) a model with one-dimensional spatial structure (the ``spatial'' model). 
For each model, we calculate the lifetime of each metastable state, \red{(which will be defined in section \ref{subsec:wm_stoch},)}  
based on the semiclassical approximation, 
and discuss the effect of spatial structure by comparing them. 
We show that the spatial \red{structure} qualitatively changes the population size dependence of the \red{lifetimes}, 
and that this change is caused by the \red{presence} of nucleation processes, i.e. 
by transitions that occur via a ``critical nucleus'' allowed only in the spatial model. 
In addition, we clarify that nucleation processes can occur only when the system's length exceeds a characteristic length. 
Although we demonstrate the spatial effect using a specific model, 
qualitatively the same result is expected to hold for general bistable evolutionary games. 

This paper is organized as follows:  
In sections \ref{sec:wmmodel} and \ref{sec:spmodel}, 
we evaluate the lifetimes of the metastable states for the well-mixed and the spatial models, respectively. 
In section \ref{sec:discussion}, by comparing the \red{lifetimes} in the two models, 
we discuss the spatial effect as well as its intuitive description. 
Section \ref{sec:summary} concludes the paper. 
The readers who are less interested in the details of the calculations may directly proceed to 
section \ref{subsec:wm_model}, \ref{subsec:sp_model}, and \ref{sec:discussion}.

\section{The well-mixed model}
\label{sec:wmmodel}

In this section, we consider the transitions between metastable states in a model without spatial \red{structure} (``well-mixed model'').
The analysis and discussion presented in this section provide an important basis for studying spatial effect in the following sections.

We define the model in section \ref{subsec:wm_model}, 
and in section \ref{subsec:wm_det}, we consider the dynamics of the expectation value of 
\red{the number (or proportion) of the strategy-A individuals},  
and confirm the bistability of the model.
We evaluate the lifetimes of the metastable states by the path integral method and the semiclassical approximation in section \ref{subsec:wm_stoch}, 
and discuss their model parameter dependence in section \ref{subsec:wm_result}. 

\subsection{Model}
\label{subsec:wm_model}

Let us assume that we have a population composed of $N (\gg 1)$ individuals, 
$n$ of which follow strategy A, and $N-n$ of which follow strategy B. 
We assume that the population is well-mixed, i.e. all the individuals play games with all the other individuals.

We describe the games performed between the individuals by a payoff matrix 
\begin{equation}
	\left(
	\begin{array}{cc}
		a & b \\
		c & d
	\end{array}
	\right),
	\label{eq:wm_model_payoffmat}
\end{equation}
which specifies payoffs of the games in the following manner: 
The game between two A individuals gives both of them a payoff $a$. 
The game between an A individual and a B individual gives them payoffs $b$ and $c$ respectively. 
The game between two B individuals gives both of them a payoff $d$. 
\red{Then, when the number of A individuals is $n$, the respective average payoff of A and B individuals are $\Pi_A(n/N)$ and $\Pi_B(n/N)$, where
\begin{eqnarray}
	\Pi_A(q) := q a  + \left( 1 - q \right) b, \qquad \Pi_B(q) := q c + \left( 1 - q \right) d. 
	\label{eq:wm_model_PF}
\end{eqnarray}
Here, $q :=n/N$ is the proportion of the A individuals in the population (note that we allowed self-interaction for simplicity). }
In this paper, we consider only a bistable ``coordination game'', i.e. impose the following conditions: 
\begin{equation}
	a > c,  \ \ b < d.
	\label{eq:wm_model_coordination}
\end{equation}
\red{The condition means that, to get higher payoff, it is always better to play the same strategy as the opponent. }

As an update rule, we adopt a ``pairwise comparison process'' \cite{TCH,TraulsenHauert}. 
In this process, two individuals, called a focal individual and a role individual, are selected randomly at a rate $\lambda$. 
The focal individual adopts the strategy of the role individual 
with probability \red{$p(\Delta \Pi) := (1+w \Delta \Pi)/2$}, 
(where $\red{\Delta \Pi := \Pi_{\rm r} - \Pi_{\rm f}}$, and 
\red{$\Pi_{\rm f}$ and $\Pi_{\rm r}$} are the payoffs of the focal and role individuals respectively). 
The parameter $w$ controls the effect of the payoffs on the spread of strategies, 
and is hence called ``the intensity of selection''. 
Thus the strategy yielding a higher payoff is more likely to be imitated by others. 
We assume $w \in [0,1]$ and $0 < a-c < 1, \  0 < d-b < 1 $ \red{to guarantee $p(\Delta \Pi) \in [0,1]$}. 

To avoid fixation, we introduce mutations (i.e. the possibility that imitation fails): 
If the role individual is A (resp. B), the focal individual adopts strategy B (resp. A) with a small probability $\mu_A$ (resp. $\mu_B$). 
We assume that the mutation rates are so small that the dynamics of the expectation value of $n$ remains bistable 
(see section \ref{subsec:wm_det}). 

The parameters used in the present paper are summarized in table \ref{tab:wm_model_param}. 
\begin{table}[tb]
	\centering
	\begin{tabular}{|c|l|} \hline
		characters & meaning \\ \hline
		$\lambda$ &  the rate at which strategy update occurs \\
		$w$ & the intensity of selection $(w \in [0,1])$ \\
		$\mu_A$ & mutation probability from A to B \\
		$\mu_B$ & mutation probability from B to A \\ 
		\red{$a,b,c,d$} & \red{elements of the payoff matrix} \\ \hline
	\end{tabular}
	\caption{Parameters used in the well-mixed model.}
	\label{tab:wm_model_param}
\end{table}

\red{
We define the transition rates for the processes, in which the number of A individuals increase or decrease by one, 
as $\lambda W_{+}(n/N)$ and $\lambda W_{-}(n/N)$, respectively, 
where $W_{\pm}$ are dimensionless transition rates. 
By adopting the update rule described above, $W_{\pm}$ are given as follows: 
\begin{eqnarray}
	\fl
	W_{+}(q) :=  (1- \mu_A)q \left(1-q \right)\frac{1 + w \left[ \Pi_A(q) - \Pi_B(q) \right] }{2} 
					+ \frac{\mu_B}{2}\left(1-q \right)^2, \label{eq:wm_model_Wp} \\
	\fl
	W_{-}(q) := (1- \mu_B)q \left(1-q \right)\frac{1 + w \left[ \Pi_B(q) - \Pi_A(q) \right] }{2}  
					+ \frac{\mu_A}{2} q^2 \label{eq:wm_model_Wm}. 
\end{eqnarray}
The system is then described by a continuous time Markov process with these transition rates. 
Note that these transition rates depend on payoff matrix elements only through $a-c$ and $d-b$ because
\begin{equation}
	\Pi_A(q) - \Pi_B(q) = (a-c+d-b)q - (d-b) \label{eq:wm_model_PFdif}.
\end{equation}
The master equation for $P(n,t)$, the probability that the number of A is  $n$ at time $t$, 
can be written as
\begin{eqnarray}
	\fl \partial_t P(n,t) = \lambda W_{+}\left(\frac{n-1}{N}\right)P(n-1,t) + \lambda W_{-}\left(\frac{n+1}{N}\right)P(n+1,t) \nonumber \\ 
		- \left[\lambda W^{+}\left(\frac{n}{N}\right) + \lambda W^{-}\left(\frac{n}{N}\right)\right] P(n,t).
	\label{eq:wm_model_mastereq}
\end{eqnarray}
}


\subsection{Deterministic dynamics of the expectation value}
\label{subsec:wm_det}

\begin{figure}[tb]
	\centering
	\includegraphics[width=10cm]{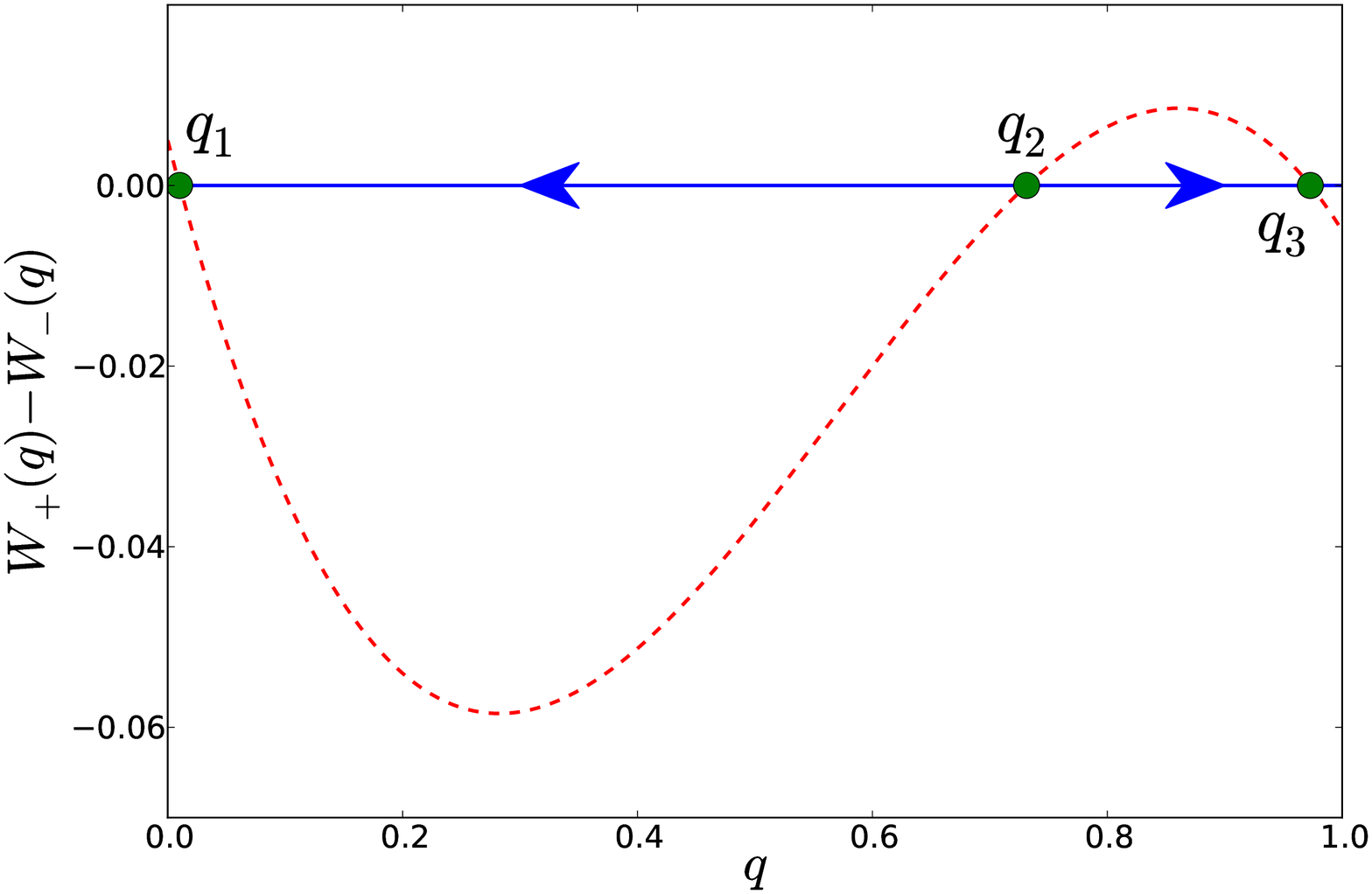}
	\caption{\red{The phase portrait of (\ref{eq:wm_det_ODE}), the deterministic dynamics of the well-mixed model. 
		Here, $q$ is the proportion of A individuals in the system.  
		The full circles denote fixed points, the dashed lines denote $W_{+}(q)-W_{-}(q)$, and arrows on the axis indicate the flow of $q$. 
		It can be seen that $q_1$ and $q_3$ are stable and that $q_2$ is unstable. Thus, the system exhibits bistability.
		Parameters: $a-c=0.4, d-b=1.0, w=0.5, \mu_A = \mu_B=0.01$}}
	\label{fig:epp}
\end{figure}

Let us first consider the dynamics of the expectation value of $n$, which we denote by $\langle n(t) \rangle$.  
In the limit $N\rightarrow \infty$, the stochastic fluctuations can be neglected and $\langle n(t) \rangle$ obeys the deterministic equation
\red{
\begin{equation}
	\frac{d}{dt} \langle n(t) \rangle = \lambda \left [ W_{+}(\langle n(t) \rangle/N) - W_{-}(\langle n(t) \rangle/N) \right].
\end{equation}
}
It is convenient to introduce the new variable $q := n/N$, which represents the proportion of A in the population,  
and rewrite the equation as, with dimensionless time $\tau := \lambda t / N$, 
\red{
\begin{eqnarray}
	\frac{d}{d\tau} \langle q(\tau) \rangle = W_{+}(\langle q(\tau) \rangle) - W_{-}(\langle q(\tau) \rangle). \label{eq:wm_det_ODE}
\end{eqnarray}
}
For the model considered here, \red{$W_{+}(q)-W_{-}(q)$} is a cubic function of $q$:
\red{
\begin{eqnarray}
	\fl W_{+}(q)-W_{-}(q) = q(1-q)\left\{ w \left(1-\frac{\mu_A + \mu_B}{2} \right)\left[(a-b-c+d)q + b-d\right] + \frac{\mu_B - \mu_A}{2}\right\}  \nonumber \\
			+ \frac{\mu_B}{2} (1-q)^2 - \frac{\mu_A}{2}q^2
			\label{eq:wm_det_cubic}
\end{eqnarray}
}
Figure \ref{fig:epp} shows a graph of \red{$W_{+}(q)-W_{-}(q)$} and the flow of $q$ determined by (\ref{eq:wm_det_ODE}). 

\red{The fixed points of the deterministic equation (\ref{eq:wm_det_ODE}) are determined by the condition $W_{+}(q)-W_{-}(q)=0$, 
i.e. the transition rates of both processes balance. }
For coordination games ($a>c, b< d$) without mutation ($\mu_A = \mu_B = 0$), it can be easily shown that the system has three fixed points: 
two stable fixed points $q_1=0$ \red{($=$ everyone adopts strategy B)} and $q_3=1$ \red{($=$ everyone adopts strategy A)} , 
and one unstable fixed point \red{$q_2=(d-b)/(a-c+d-b)$ ($=$ mixture of strategy A and B)}.
The positions of the fixed points shift slightly when small mutations are introduced. 
However, the system remains bistable, i.e. it has two stable fixed points ($q_1$ and $q_3$) and one unstable fixed point ($q_2$) in the range $[0,1]$ 
(see figure \ref{fig:epp}). 
Hereafter, we assume that the mutation rates $\mu_A$ and $\mu_B$ are sufficiently small so that the system remains bistable. 


\subsection{Transitions between metastable states}
\label{subsec:wm_stoch}

As discussed above, the expectation value of $n$ exhibits bistability. 
In addition, as can be seen from the definition of the model, 
\red{fixation to one strategy is impossible} because of mutations. 
Thus, although the system remains at states $q_1$ or $q_3$ for an extremely long time, 
large fluctuations can occasionally carry the system from one state to the other. 
Therefore, states $q_1$ and $q_3$ are metastable and have a long but finite lifetime. 

\red{
To demonstrate this feature, we performed a Monte Carlo simulation of the stochastic process described by the master equation (\ref{eq:wm_model_mastereq}) 
based on the Gillespie algorithm \cite{Gillespie}. 
\begin{figure}[tb]
	\centering
	\includegraphics[width=12cm]{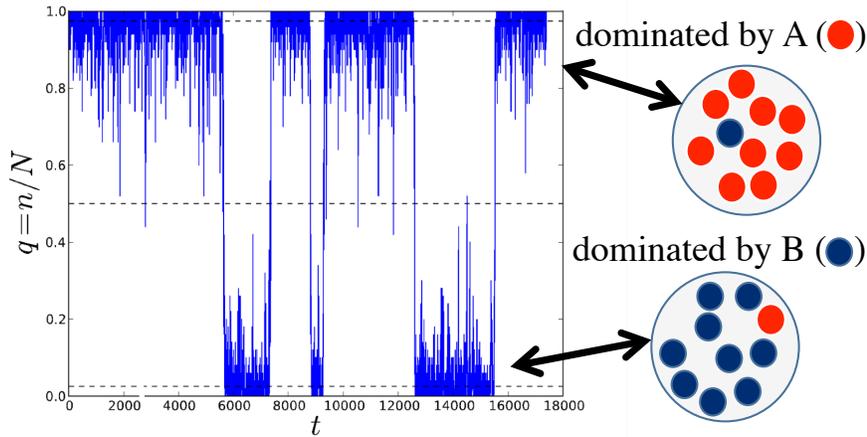}
	\caption{\red{Time evolution of $q:=n/N$ (the proportion of A individuals in the population) obtained by a Monte Carlo simulation of 
		(\ref{eq:wm_model_mastereq}), the stochastic dynamics of the well-mixed model. 
		Dashed lines denote $q_1, q_2$ and $q_3$. 
		The system clearly exhibits metastability: though the system stays around the metastable states (B-dominant state $q_1$ or A-dominant state $q_3$) for a long time, 
		it occasionally undergoes transition between these states. }
		\red{Parameters: $N=50, a-c=d-b=0.5, w= 0.4, \mu_A = \mu_B = 0.01$. 
		For these parameters, the positions of the metastable states are $q_1 \simeq 0.0259$, $q_3 \simeq 0.974$. } }
	\label{fig:metastable}
\end{figure}
We show an example of stochastic time evolution of the number of A individuals for $N=50$ in figure \ref{fig:metastable}, 
which shows that while the system stays around the two stable fixed points ($q_1$ and $q_3$) for a long time, 
there are rare transitions between these two metastable states due to stochastic fluctuations. 
These transitions are expected to be very rare when $N$ is large, because fluctuations around metastable states are suppressed. 
}

In this paper, we evaluate the \red{lifetimes} of these metastable states, 
\red{which are defined to be the mean waiting times until the system escapes from the given metastable state and undergoes transition to the other metastable state}. 
In the following, we \red{mainly focus on} the lifetime of the metastable state $q_3$, 
because the lifetime of the metastable state $q_1$ can be calculated in the same manner. 


\subsubsection{Path integral expression}
\label{subsubsec:wm_stoch_PI}

\red{The lifetime of the metastable state $q_3$ is calculated from }
$P_{q_3,\rm{wm}}(T)$, the probability that the system stays around $q_3$ from $\tau = 0$ to $\tau = T (\gg 1 )$
without ever visiting the other metastable state $q_1$. 
$P_{q_3,\rm{wm}}(T)$ decays exponentially with $T$, and the inverse of the decay rate gives the lifetime of $q_3$. 

\red{
$P_{q_3,\rm{wm}}(T)$ can be approximated by the probability that $q(T) = q_3$ and $q(\tau) \neq q_1 (\forall \tau \in [0,T] )$ given that $q(0) = q_3$. 
We express the latter probability using the path integral formulation of stochastic processes \cite{AltlandSimons,LefevreBiroli}, 
in which the probability of given paths can be expressed as a summation, with some weight, over the paths.  
With this technique, we obtain (see \ref{sec:SPandPI} for the derivation)
}
\begin{eqnarray}
	P_{q_3,\rm{wm}}(T) &\simeq& \int'_{q(0) = q_3, q(T) = q_3 } \mathcal{D} q \mathcal{D} p \  e^{-N S[q(\cdot),p(\cdot)]} 
		\label{eq:wm_Stoch_PI_prob}, \\
	S[q(\cdot),p(\cdot)] &:=& \int_{0}^{T} d\tau  [ p(\tau) \partial_{\tau} q(\tau) - H_0(q(\tau),p(\tau)) ], \\
	H_0(q,p) &:=& ( e^p - 1 ) W_{+}(q) + (e^{-p} - 1)W_{-}(q) \label{eq:wm_Stoch_PI_Hdef}, 
\end{eqnarray}
where $W_{\pm}$ are given by (\ref{eq:wm_model_Wp}) and (\ref{eq:wm_model_Wm}), 
$p$ is a new variable conjugate to $q$, 
and $\int'_{q(0) = q_3, q(T) = q_3 } \mathcal{D}q  \mathcal{D}p$ represents the (restricted) summation over all the paths $(q,p)$ 
satisfying $q(0) = q(T) = q_3$ and $q(\tau) \neq q_1 (\forall \tau \in [0,T] )$. 
\red{
The meaning of this expression is that the probability density of each path $q(\cdot),p(\cdot)$ is $\exp(-NS[q(\cdot),p(\cdot)])$, 
and that the desired probability is a weighted sum over all paths satisfying the conditions. 
Note that the present formalism is an analog of the path integral in quantum mechanics. 
From this analogy, $S$ and $H_0$ are called an action and a Hamiltonian, respectively. 
The path integral expression is useful for the analysis of $N \gg 1$ cases as discussed in the next section. 
}


\subsubsection{Semiclassical approximation}
\label{subsubsec:wm_stoch_SC}

We will now evaluate the path integral expression (\ref{eq:wm_Stoch_PI_prob}) \red{under the assumption $N \gg 1$}. 
\red{We adopt a semiclassical approximation, 
in which the path integral is approximated by contribution from the stationary paths $(q,p)$ of action $S$ (i.e. $\delta S[q,p] = 0$) 
and the fluctuations around them
because, when $N$ is large, only the stationary paths contribute dominantly to the path integral. 
This approximation is an extension of the steepest descent method in evaluating integrals, and an analog of the 
WKB (or semiclassical) approximation in quantum mechanics. 
It is known that the semiclassical approximation is suitable for treating rare events \cite{AltlandSimons}. }

\begin{figure}[tb]
	\centering
	\includegraphics[width=15cm]{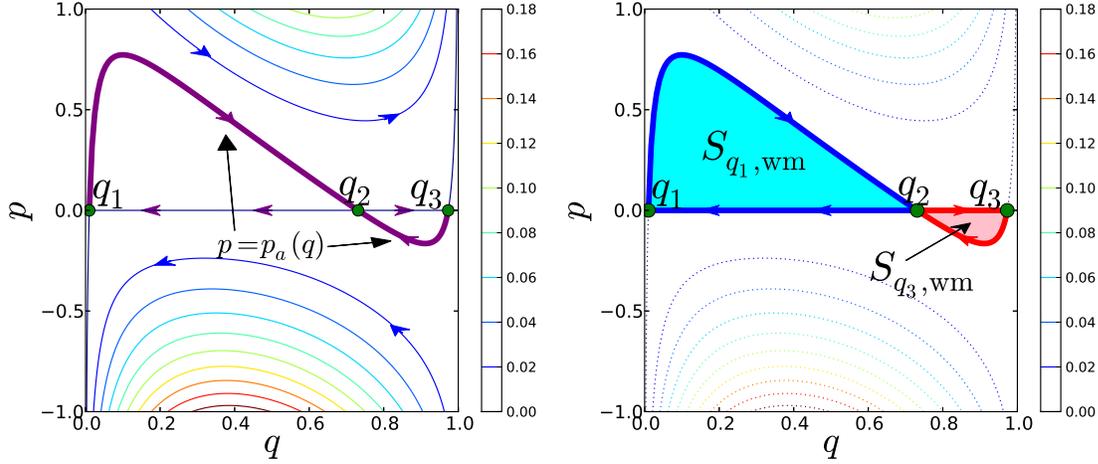}
	\caption{\red{Phase portraits of the equations of motion (\ref{eq:wm_Stoch_SC_eqmq}) and (\ref{eq:wm_Stoch_SC_eqmp}). 
		$q$ is the fraction of $A$ individuals in the system, and $p$ is a variable conjugate to $q$. 
		The arrows indicate the flows of (\ref{eq:wm_Stoch_SC_eqmq}) and (\ref{eq:wm_Stoch_SC_eqmp}), }
		\red{three} green circles denote fixed points, and \red{colored curves} show the contours of the Hamiltonian. 
		\red{Left: Bold lines denote activation trajectories $p_a$ (see (\ref{eq:wm_Stoch_SC_activationtrajectory}))
		Right: Bold lines denote closed trajectories that determine the lifetimes of the metastable states. 
		The red trajectory (right) determines the lifetime of $q_3$, while the blue trajectory (left) determines the lifetime of $q_1$. 
		The area shaded in red (right) is equal to $S_{q_3, {\rm wm}}$ (see (\ref{eq:wm_Stoch_SC_Sbounce})).
		Parameters: $a-c=0.4, d-b=1.0, w=0.5, \mu_A = \mu_B=0.01$}}
	\label{fig:scpp}
\end{figure}

The stationary condition $\delta S = 0$ yields \red{differential equations}: 
\begin{eqnarray}
	\frac{dq}{d\tau} &=& \frac{\partial H_0}{\partial p}  = e^p \ W_{+}(q) - e^{-p} \ W_{-}(q), \label{eq:wm_Stoch_SC_eqmq} \\
	\frac{dp}{d\tau} &=& -\frac{\partial H_0}{\partial q} =- (e^p -1)W'_{+}(q) - (e^{-p}-1)W'_{-}(q), \label{eq:wm_Stoch_SC_eqmp}
\end{eqnarray}
(where $W'_{\pm}(q) := dW_{\pm}(q)/dq$)
subjected to the boundary conditions $q(0) = q(T) = q_3$ and the condition $q(\tau) \neq q_1 (\forall \tau \in [0,T] )$. 
\red{Because the equations have the same form as Hamilton's equations of motion in analytical mechanics, }
the solutions $(q,p)$ satisfying these \red{differential equations} are called ``classical'' trajectories in the following discussion. 
Figure \ref{fig:scpp} shows phase portraits of \red{the flow of} these equations. 
\red{Note that $H_0(q,p)$ is conserved along each trajectory. }

\red{
There are two important types of trajectories. 
The first one is the trajectories on the horizontal line of $p=0$. 
These trajectories represent the dynamics of the expectation value of $q$, 
because the equation of motion (\ref{eq:wm_Stoch_SC_eqmq}) for $p=0$ coincides with the equation of $\langle q \rangle$ (\ref{eq:wm_det_ODE}). 
Note that $H_0(q,p) = 0$ for these trajectories.  
}

\red{
The second one is the trajectories shown by bold lines in the left panel of figure \ref{fig:scpp}, 
connecting stable fixed points ($(q_1,0)$ and $(q_3,0)$) and unstable fixed point ($(q_2,0)$). 
These two trajectories are called ``activation trajectories'' \cite{EscuderoKamenev,AssafMeerson2010}. 
Because the Hamiltonian is always constant on the connected trajectories, $H_0(q,p) = 0$ holds on the activation trajectories. 
Hence, the shape of the activation trajectories can be collectively expressed as $p = p_a(q)$, where
\begin{equation}
	p_a(q) := - \log \left[ \frac{W_{+}(q)}{W_{-}(q)} \right], 
	\label{eq:wm_Stoch_SC_activationtrajectory}
\end{equation}
which can be obtained from the condition $H(q,p_a(q)) = 0$ (see (\ref{eq:wm_Stoch_PI_Hdef})).
}

\red{
In the semiclassical approximation of $P_{q_3, {\rm wm}}$, classical trajectories should be properly chosen 
so that the conditions $q(0) = q(T) = q_3$ and $q(\tau) \neq q_1 \ (\forall \tau \in [0,T])$ are satisfied. 
Therefore, the classical trajectories relevant for the evaluation of $P_{q_3, {\rm wm}}$ are restricted to 
(i) a trivial solution $q(\tau) \equiv q_3$, and 
(ii) nontrivial solutions circulating on the closed trajectory shown in the right panel of figure \ref{fig:scpp} (the red closed trajectory), 
i.e. the closed trajectory composed of an activation trajectory and a $p=0$ trajectory. 
Note that only these trajectories spend significant amount of  time in going around and 
can satisfy the boundary conditions $q(0) = q(T) = q_3$ for $T \gg 1$. 
These solutions are called ``bounce solutions'' \cite{AltlandSimons}.  
}

\red{
The action of the trivial solution is zero, whereas the action of nontrivial solutions are given by $n S_{q_3, {\rm wm}}$, where
$n (= 1,2,3,\cdots)$ is the number of rotation and 
$S_{q_3, {\rm wm}}$ is the action per one cyclic motion on the closed trajectory described above.
Using the zero-energy condition $H_0(q,p)=0$ and the expression for the activation trajectories $p_a(q)$, 
$S_{q_3, {\rm wm}}$ can be expressed as
\begin{eqnarray}
	S_{q_3, {\rm wm}} = \int_{q_3}^{q_2} p_a(q) dq \label{eq:wm_Stoch_SC_Sbounce} ,
\end{eqnarray}
which is equal to the area shaded in red in the right panel of figure \ref{fig:scpp}.  
}
Summing up all the contributions from the bounce solutions, we obtain
\begin{eqnarray}
	P_{q_3,\rm{wm}}(T) 
	&\simeq& \sum_{n=0}^{\infty} \int_{0}^{t_2} dt_1 \cdots \int_{0}^{T} dt_n \left( -K e^{-NS_{q_3, {\rm wm}}} \right)^n \nonumber \\
	&=&  \exp \left( - K e^{-NS_{q_3, {\rm wm}} } T  \right)	\label{eq:wm_Stoch_SC_Psurvive}.
\end{eqnarray}
Here, the prefactor $K$ is a positive value determined by Gaussian integrals around bounce solutions. 

Thus, the lifetime of the metastable state $q_3$, $\tau_{q_3, {\rm wm}}$, can be expressed as 
\begin{equation}
	\tau_{q_3, {\rm wm}} \simeq \frac{1}{K} \exp(N S_{q_3, {\rm wm}}).
	\label{eq:wm_Stoch_SC_lifetime}
\end{equation}
The lifetime of the other metastable state $q_1$ can be evaluated in the same manner, 
except that the other \red{closed trajectory (the blue trajectory in the right panel of figure \ref{fig:scpp})} should be used: 
\red{
\begin{eqnarray}
	\tau_{q_1, {\rm wm}} \simeq \frac{1}{K'} \exp\left( N S_{q_1, {\rm wm}}\right), \label{eq:wm_Stoch_SC_lifetime2} \\
	S_{q_1, {\rm wm}} := \int_{q_1}^{q_2} p_a(q) dq \label{eq:wm_Stoch_SC_Sbounce2}, 
\end{eqnarray}
where the prefactor $K'$ is a positive quantity (determined by Gaussian integrals around bounce solutions around $q_1$). 
} 

It can be seen from (\ref{eq:wm_Stoch_SC_lifetime}) \red{and (\ref{eq:wm_Stoch_SC_lifetime2})} 
that the lifetimes are extremely long ($\tau_{q_i, {\rm wm}} \gg 1$)
under the condition $N \gg 1$ assumed in this paper. 
The lifetimes strongly depend on the action \red{$S_{q_i, {\rm wm}}$} because of the presence of a large factor $N$ in the exponent. 
In this paper, we focus on the exponent of the lifetimes, and neglect the weak parameter dependence of the prefactors $K$ and $K'$.


\subsection{Result}
\label{subsec:wm_result}

In this section, we briefly summarize parameter dependence of the lifetime $\tau_{q_i, {\rm wm}} \ (i \in \left\{1,3\right\})$. 

\subsubsection{$N$ dependence}
\label{subsubsec:wm_result_Ndep}

Because action $S_{q_i, {\rm wm}}$ is positive and independent of $N$, 
the lifetimes $\tau_{q_i, {\rm wm}} \propto \exp(N S_{q_i, {\rm wm}}) \ (i \in \left\{ 1,3 \right\})$ increases exponentially with $N$. 
This result is consistent with the previous one, which showed that the fixation probability in the coordination game decays exponentially with $N$ \cite{AntalScheuring}. 
This kind of exponential dependence of some ``lifetimes'' on population size is known for various models in which the population is well-mixed. 
Examples include ecological models (the mean time to the extinction of a population) \cite{AssafMeerson2010,OvaskainenMeerson,DoeringEtal}, 
evolutionary games (the mean time to fixation in anti-coordination games) \cite{AssafMobilia} and general reaction models \cite{EscuderoKamenev}.

\subsubsection{$w$ dependence}
\label{subsubsec:wm_result_Wdep}

\begin{figure}[tb]
	\centering
	\includegraphics[width=8cm]{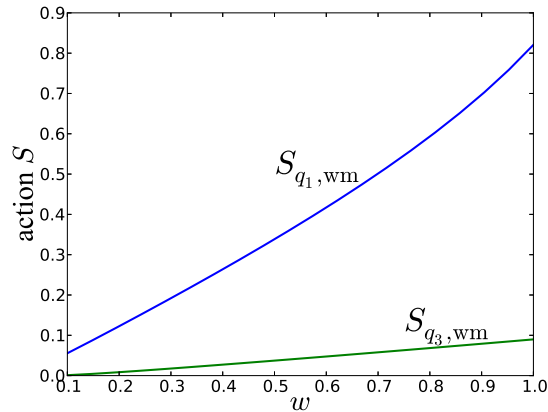}
	\caption{$w$ dependence of $S_{q_1, {\rm wm}}$ and $S_{q_3, {\rm wm}}$. 
		\red{$S_{q_i, \rm{wm}}$ increases with $w$. }
		Parameters: $a-c=0.4, d-b = 1.0, \mu_A = \mu_B = 0.005$.}
	\label{fig:wmWdep}
\end{figure}

Figure \ref{fig:wmWdep} shows $S_{q_1, {\rm wm}}$ and $S_{q_3, {\rm wm}}$ as the functions of $w$ (the intensity of selection). 
As can be seen from the figure, the action grows with $w$. 
This result indicates that, when $N \gg 1$, the lifetimes $\tau_{q_i, {\rm wm}} \propto \exp(N S_{q_i, {\rm wm}}) \ (i \in \left\{ 1,3 \right\})$ 
increase rapidly as natural selection becomes strong.

\subsubsection{Payoff matrix dependence}
\label{subsubsec:wm_result_PFdep}

\begin{figure}[tb]
	\centering
	\includegraphics[width=15cm]{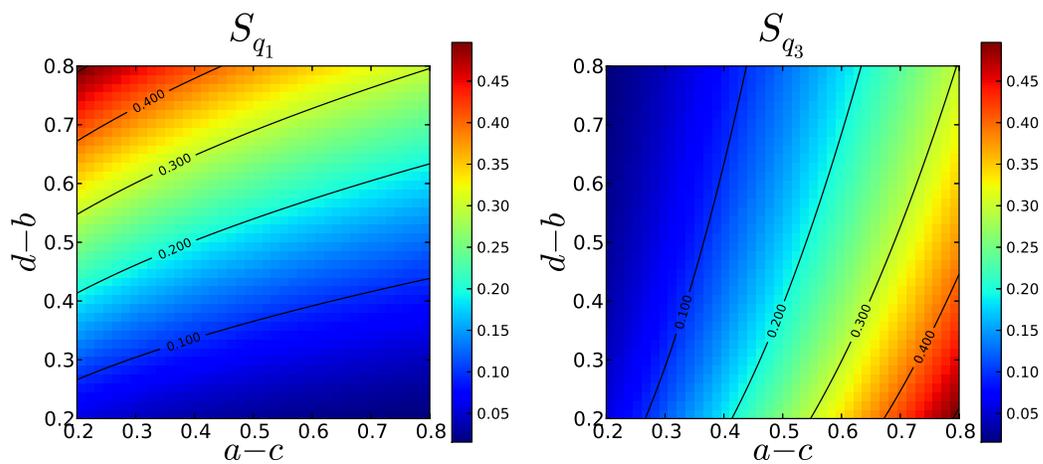}
	\caption{The payoff matrix dependence of $S_{q_1, {\rm wm}}$(left) and $S_{q_3, {\rm wm}}$(right). Parameters: $w = 0.8, \mu_A = \mu_B = 0.005$}
	\label{fig:wmPFdep}
\end{figure}

As can be seen from (\ref{eq:wm_model_Wp})-(\ref{eq:wm_model_PFdif}), 
the stochastic dynamics of the evolutionary game considered here depends on payoff matrix elements only through $a-c$ and $d-b$. 
The dependence of $S_{q_1, {\rm wm}}$ and $S_{q_3, {\rm wm}}$ on these two parameters is shown in figure \ref{fig:wmPFdep} 
(only the region $0.2 < a-c < 0.8$ and $0.2 < d-b <0.8 $ is plotted because outside this region there are parameter sets for which the system is not bistable). 

This dependence can be intuitively explained in the following manner: 
The larger $a-c$ is, the more advantageous A individuals is in population dominated by A, compared with B individuals in the same population, 
and the longer the lifetime of $q_3$. 
Almost the same discussion applies to $d-b$ and the lifetime of $q_1$. 
Note that when $\mu_A = \mu_B$, a symmetry relation $S_{q_1, {\rm wm}}(a-c,d-b) = S_{q_3, {\rm wm}}(d-b,a-c)$ holds. 

\section{Spatial model}
\label{sec:spmodel}

On the basis of the analysis presented in the previous section,  
in this section we examine the transition between metastable states in a model with spatial structure (``spatial model'').
The discussion proceeds almost parallel to that of the previous section. 
The model is defined in section \ref{subsec:sp_model}. 
In section \ref{subsec:sp_det}, we analyze the behaviour of the expectation value, and determine steady states and their stability. 
In section \ref{subsec:sp_stoch}, we evaluate the lifetimes of metastable states using the path integral expression and the semiclassical approximation. 
In section \ref{subsec:sp_result}, we show how these lifetimes depend on the system size.  


\subsection{Model}
\label{subsec:sp_model}

\begin{figure}[tb]
	\centering
	\includegraphics[angle=-90,width = 10cm]{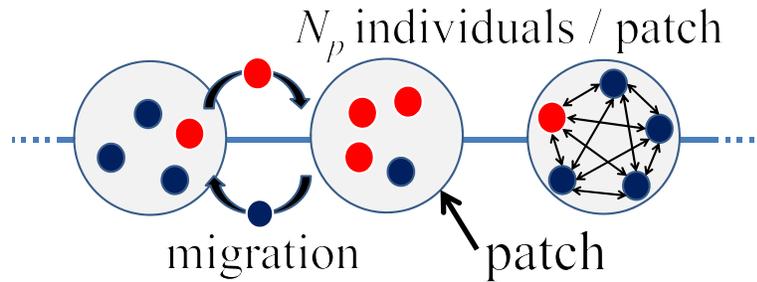}
	\caption{A schematic diagram of the spatial model. 
		\red{The small circles denote individuals (Red: A individuals, Blue: B individuals). Individuals play games and reproduce inside each patch, and can move between neighbouring patches. }}
	\label{fig:sp_patch}
\end{figure}

We consider a one-dimensional array of $M$ patches. 
Let $l$ be the separation between neighbouring patches and $\tilde{L} := Ml$ be the length of the system. 
We impose a periodic boundary condition. 
In each patch, there are $N_p (\gg 1)$ individuals, who take either strategy A or strategy B, 
and change their strategies according to the same evolutionary game as described in section \ref{sec:wmmodel}. 
We further assume that the individuals can move between neighbouring patches. 
This migration process is modeled as the ``swapping'' of individuals so that the number of individuals per patch is conserved. 

Let $n_i$ be the number of A individuals in the $i$th patch ($i \in \left\{ 1, 2, \cdots, M \right\}$). 
The parameters of the evolutionary rule in each patch are the same as those described in section \ref{sec:wmmodel} 
(see table \ref{tab:wm_model_param}). 
\red{The transition rates of $n_i$ due to strategy update can be written as $\lambda W_{+} (n_i /N_p)$ and $\lambda W_{-}(n_i /N_p)$, 
where $W_{\pm}$ are given in (\ref{eq:wm_model_Wp}) and (\ref{eq:wm_model_Wm}), respectively.}

Migration (swapping) processes are defined as follows: 
A pair of neighbouring patches, $i$ and $j$, is randomly chosen at a rate $\sigma$. 
One individual is randomly picked up from each patch and they are swapped. 
\red{The rate of the process $(n_i, n_j) \rightarrow (n_i -1, n_j +1)$ due to swapping is given by $\sigma W_{\rm m}(n_j/N_p, n_j/N_p)$, where
\begin{equation}
	W_{\rm m}(q_i,q_j) := q_i (1-q_j). 
\end{equation}
}

With these transition rates, 
the master equation for the probability distribution $P(\boldsymbol{n}, t)$ on the population configuration $\boldsymbol{n} = (n_1, n_2, \cdots, n_M)$ is written as
\begin{eqnarray}
	\fl
	\partial_t P(\boldsymbol{n}, t) &=& 
		\sum_{i=1}^{M} \left\{ \lambda W_{+}\left( \frac{n_i-1}{N_p} \right)P(\boldsymbol{n} - \boldsymbol{e}_i, t) 
				+ \lambda W_{-}\left(\frac{n_i+1}{N_p}\right)P(\boldsymbol{n} + \boldsymbol{e}_i, t) \right. \nonumber \\
	\fl		&& \qquad \left. - \left[ \lambda W_{+}\left(\frac{n_i}{N_p}\right) + \lambda W_{-}\left(\frac{n_i}{N_p}\right) \right] P(\boldsymbol{n}, t) \right\} \nonumber \\
	\fl 	&& + \sum_{\langle i, j \rangle} \left[ \sigma W_{\rm m} \left(\frac{n_i+1}{N_p}, \frac{n_j-1}{N_p}\right) P(\boldsymbol{n} + \boldsymbol{e}_i - \boldsymbol{e}_j, t) \right. \nonumber \\
	\fl		&& \qquad \left. - \sigma W_{\rm m}\left(\frac{n_i}{N_p}, \frac{n_j}{N_p} \right) P(\boldsymbol{n},t) \right],
		\label{eq:sp_model_mastereq}
\end{eqnarray}
where $\langle i,j \rangle$ indicates neighbouring patches, and 
\begin{equation}
	\boldsymbol{e}_i := (0, \cdots, 0, \stackrel{i}{\breve{1}}, 0, \cdots, 0). 
\end{equation}


\subsection{Dynamics of the expectation value}
\label{subsec:sp_det}

In this section, we derive a differential equation for the expectation values of $n_i$, 
and examine its steady solutions and their stability, 
\red{which play an important role in understanding the transitions between metastable states in the spatial model. }


\subsubsection{Deterministic equations}
\label{subsubsec:sp_det_eq}

Let $\langle n_i \rangle$ be the expectation value of $n_i$. 
Then, $\langle n_i \rangle$ ($i \in \left\{ 1,2, \cdots, M \right\}$) obey 
\red{
\begin{eqnarray}
	\fl
	\frac{d \langle n_i \rangle}{dt} &= \lambda W_{+}(\langle n_i\rangle / N_p)  - \lambda W_{-}(\langle n_i \rangle / N_p) \nonumber \\
	\fl &{}
			+  \sigma W_{\rm m}(\langle n_{i+1} \rangle /N_p, \langle n_i \rangle /N_p) + \sigma W_{\rm m}(\langle n_{i-1}\rangle /N_p, \langle n_i \rangle /N_p)   \nonumber \\
	\fl &{} \qquad  - \sigma W_{\rm m}(\langle n_i \rangle /N_p, \langle n_{i+1}\rangle /N_p ) - \sigma W_{\rm m}(\langle n_i \rangle /N_p, \langle n_{i-1}\rangle /N_p ). 
\end{eqnarray}
}
By introducing new variables $q_i := n_i/N_p$ \red{(the proportion of A individuals in the $i$th patch)} and a rescaled time $\tau := \lambda t/N_p$, 
we obtain
\red{
\begin{eqnarray}
	\frac{d \langle q_i \rangle}{d\tau} &=&  W_{+}(\langle q_i \rangle)  - W_{-}(\langle q_i \rangle)
		+ \frac{\sigma}{\lambda} \left( \langle q_{i+1} \rangle + \langle q_{i-1} \rangle - 2 \langle q_{i} \rangle \right) \label{eq:sp_det_eq_rateeq}. 
\end{eqnarray}
}

In this paper, we assume $\sigma \gg \lambda$, i.e. migration processes occur sufficiently faster than \red{strategy update processes}. 
Under this assumption, $\langle q_i \rangle$ changes smoothly as a function of $i$, 
and can be expressed by a function of a continuum spatial degree of freedom $x$: 
\begin{equation}
	q(x,\tau) := \langle q_{\frac{x}{l}}(\tau) \rangle  \qquad (x \in [0, \tilde{L}]).  
\end{equation}
The second term on the right hand side of (\ref{eq:sp_det_eq_rateeq}) is approximated by $D\partial^2 q / \partial \xi^2$, 
where $D := \sigma l^2 / \lambda$ is a diffusion constant. 
By introducing the dimensionless space variable $\xi := x/ \sqrt{D}$, 
we arrive at the following reaction diffusion equation: 
\begin{eqnarray}
	\frac{\partial q(\xi,\tau)}{\partial \tau} = \frac{\partial ^2 q(\xi,\tau)}{\partial {\xi}^2} + W_{+}(q(\xi,\tau)) - W_{-}(q(\xi,\tau))  \qquad
		\left(\xi \in [0,L] \right), 
	\label{eq:sp_det_eq_PDE}
\end{eqnarray}
where $L := \tilde{L}/\sqrt{D} $ \red{is the rescaled system size}.


\subsubsection{Steady solutions}
\label{subsubsec:sp_det_steady}

During the analysis of the well-mixed model discussed in section \ref{sec:wmmodel}, 
fixed points and their stability played important roles. 
In the spatial model, steady solutions will play similar roles. 

\begin{figure}[tb]
	\centering
	\includegraphics[width=15cm]{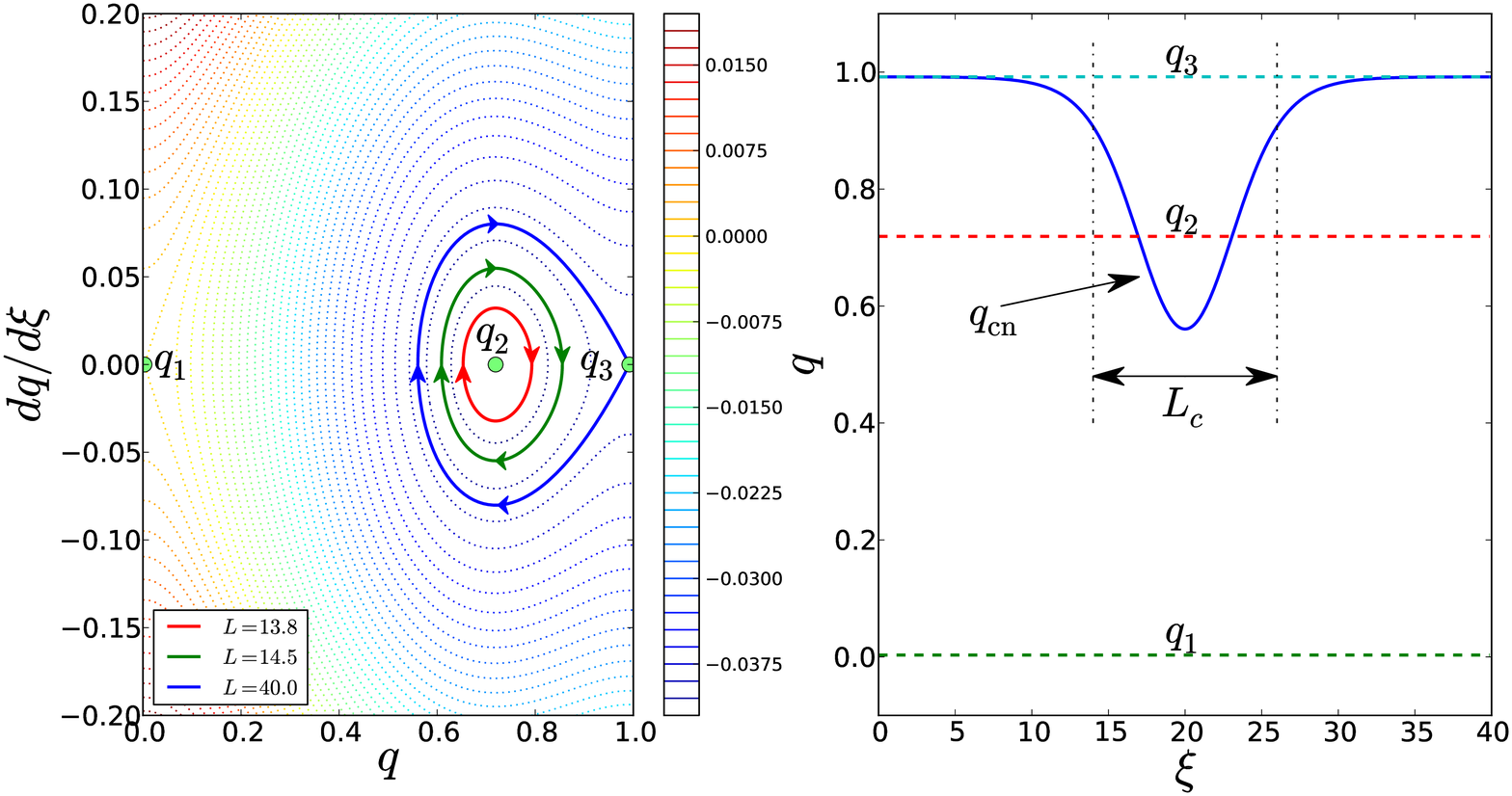}
	\caption{
		Steady solutions of (\ref{eq:sp_det_eq_PDE}). 
		Left: A ``phase portrait'' of (\ref{eq:sp_det_ss_ODE}). The solid lines represent critical nuclei. 
		The color bar denotes the value of $E$ defined in (\ref{eq:sp_det_ss_energy}).
		Right: A critical nucleus ($L = 40$). 
		$L_c$ is the typical scale of a critical nucleus. 
		Parameters: $a-c=0.4, d-b=1.0, w=0.8,\mu_A=\mu_B=0.005$ ($L_c \simeq 13.5$ and $V(q_1) > V(q_3)$ in this case). 
		The critical nuclei were numerically calculated by applying Newton's method to a discretized version of (\ref{eq:sp_det_ss_ODE}).}
	\label{fig:cnuc}
\end{figure}

Let $q_s(\cdot)$ be a steady solution of (\ref{eq:sp_det_eq_PDE}). 
$q_s$ obeys an ordinary differential equation
\red{
\begin{equation}
	\frac{d^2 q_s(\xi)}{d {\xi}^2} + W_{+}(q_s(\xi)) - W_{-}(q_s(\xi)) = 0.
	\label{eq:sp_det_ss_ODE}
\end{equation}
}
Integrating this equation yields 
\red{
\begin{eqnarray}
	\frac{1}{2} \left( \frac{d q_s(\xi)}{d\xi} \right) ^2 + V(q_s(\xi)) = {\rm const.} =:E \label{eq:sp_det_ss_energy}, \\
	V(q) := \int_{0}^{q} \left[ W_{+}(q') - W_{+}(q') \right] dq'. \label{eq:sp_det_ss_potential}
\end{eqnarray}
}
By regarding $\xi$ as the ``time'', we can describe $q_s$ as a coordinate of a particle moving in one dimension, subjected to potential $V$. 
Hence, the trajectories of $q_s$ on the $(q,dq/d\xi)$ plane can be expressed by the contours of ``energy'' $E$ defined in (\ref{eq:sp_det_ss_energy}) 
(dotted lines in the left panel of figure \ref{fig:cnuc}). 
Because of the periodic boundary condition $q_s(\xi+L) =q_s(\xi)$, $q_s$ must be a closed orbit or a point on the $(q,dq/d\xi)$ plane. 
Therefore, the steady solutions must be
\begin{description}
	\item[(a)] uniform solutions:  $q_s(\xi) \equiv q_i \ (i \in \left\{ 1,2,3 \right\} )$ \red{(recall that $q_i$ are the solutions of $W_{+}(q)-W_{-}(q)=0$ defined in section \ref{subsec:wm_det})} or
	\item[(b)] non-uniform periodic solutions: closed orbits surrounding $q_2$. 
\end{description}

First, we consider the uniform solutions, which always exist regardless of system size $L$. 
\red{Because $W'_{+}(q) - W'_{-}(q)$ is negative for $q_1, q_3$ and positive for $q_2$ (see figure \ref{fig:epp}), }
two solutions, $q_s(\xi) \equiv q_1$ and $q_s(\xi) \equiv q_3$, are linearly stable, and $q_s(\xi) \equiv q_2$ is linearly unstable; 
the former solutions correspond to the two metastable states, 
whereas the latter solution corresponds to the ``marginal'' state located at the boundary $q_2$, 
below which the system goes to $q_1$ and above which the system goes to $q_3$. 

Next, we examine the properties of the non-uniform solutions. 
The forms of these solutions depend on the system size $L$, 
because the period of the solution must coincide with $L$: 
\begin{equation}
	2\int_{q_{\rm min}}^{q_{\rm max}} \frac{dq}{ \sqrt{2(E-V(q))} }  =L.
\end{equation}
Here, $q_{\rm min}$ and $q_{\rm max}$ are the minimum and the maximum of $q$ in the trajectory, 
which is determined by $E$. 
The left panel of figure \ref{fig:cnuc} show the phase portrait for the case of $V(q_1) > V(q_3)$, 
where the non-uniform solution in the limit of $L \rightarrow \infty$ corresponds to the 
homoclinic orbit starting from $q_3$ and ending at $q_3$ (the blue line in the figure). 
The right panel of figure \ref{fig:cnuc} shows the spatial profile of a non-uniform solution for the same parameter set as the left panel. 
As can be seen from the figure, the solution represents a ``nucleus'' of B individuals surrounded by a region dominated by A individuals. 
Note that for the opposite case (i.e. $V(q_1) < V(q_3)$), \red{the form of} the non-uniform solution becomes ``upside down'' of the right panel of figure \ref{fig:cnuc}, 
i.e. it represents a nucleus of A individuals surrounded by a region dominated by B individuals. 

\red{
It can be shown that the obtained non-uniform solutions are unstable, 
i.e. even an infinitesimally small perturbation drives the system away from the non-uniform solutions and toward the stable states $q_1$ or $q_3$ (see \ref{sec:Stability_cn}). 
Because of such ``critical'' behaviour, these non-uniform steady solutions are called ``critical nuclei'' \cite{MeersonSasorov}, and we denote them by $q_{\rm cn}$. 
}

As $L$ decreases, both $q_{\rm min}$ and $q_{\rm max}$ approach $q_2$, 
and at a critical length $L_c$, they coalesce with $q_2$. 
This indicates that critical nuclei do not exist when $L$ is smaller than $L_c$, 
which is calculated, \red{by linearizing (\ref{eq:sp_det_ss_ODE}) around $q_2$,} as
\begin{equation}
	L_c = \frac{2\pi}{\sqrt{W'_{+}(q_2)- W'_{-}(q_2)}}. 
	\label{eq:sp_det_ss_Lc}
\end{equation} 
$L_c$ gives the characteristic length scale of the critical nuclei and plays an important role in considering the spatial effect on 
the transitions between metastable states as discussed later in section \ref{sec:discussion}.


\subsection{Transitions between metastable states}
\label{subsec:sp_stoch}

As shown in the previous section, the spatial model considered deterministically has two stable steady solutions, i.e.
the uniform solutions $q_s(\xi) \equiv q_1$ and $q_s(\xi) \equiv q_3$, 
which we henceforth simply call $q_1$ and $q_3$, respectively. 
If stochasticity is taken into account, these solutions correspond to metastable states: 
although the system stays at $q_1$ or $q_3$ for an extremely long time, 
it occasionally undergoes transitions from one state to the other as a result of large stochastic fluctuations. 
In this section, we evaluate the lifetimes of these metastable states. 
We show the calculation of the lifetime of $q_3$ (the lifetime of $q_1$ can be calculated in the same manner).


\subsubsection{Path integral expression}
\label{subsubsec:sp_stoch_PI}

To evaluate the lifetime of a metastable state $q_3$, 
we calculate $P_{q_3, \rm{sp}}(T)$, the probability that the system stays around $q_3$ from $\tau=0$ to $\tau=T (\gg 1)$
without ever visiting the other metastable state $q_1$. 
$P_{q_3, \rm{sp}}(T)$ decays exponentially with $T$, and the inverse of the decay rate gives the lifetime of $q_3$. 

The path integral formalism, which was applied to the well-mixed model in section \ref{sec:wmmodel}, 
is also applicable to the spatial model (see \ref{sec:SPandPI} for the detail). 
\red{$P_{q_3, {\rm sp}}$ can be approximated by the probability that $q(\cdot,T) = q_3$ and $q(\cdot,\tau) \neq q_1 \ (\forall \tau \in [0,T])$ 
given that $q(\cdot, 0) = q_3$. }
By adopting continuum description assuming $\sigma \gg \lambda$ 
and using a dimensionless spatial coordinate $\xi := jl/\sqrt{D} \ \ ( j \in \left\{ 1,2, \cdots, M \right\} )$ 
and the rescaled system size $L := \tilde{L}/\sqrt{D}$,
we obtain
\begin{eqnarray}
	\fl
	P_{q_3, \rm{sp}}(T) \simeq \int'_{q(\cdot, 0) = q_3, q(\cdot, T) = q_3} 
		\mathcal{D} q\mathcal{D}p \exp \left( -N_p \sqrt{\frac{\sigma}{\lambda}} S[q(\cdot,\cdot), p(\cdot,\cdot)] \right) 
		\label{eq:sp_stoch_PI_prob}, \\
	\fl
	S[q(\cdot,\cdot),p(\cdot,\cdot)] := \int_{0}^{T} d\tau \int_{0}^{L} d\xi
								\left\{ p(\xi,\tau) \partial_{\tau}q(\xi,\tau) - h [q(\cdot,\cdot),p(\cdot,\cdot)](\xi,\tau) \right\} 
		\label{eq:sp_stoch_PI_ScontS}, \\
	\fl
	h[q(\cdot,\cdot),p(\cdot,\cdot)](\xi,\tau) :=
			h_0(q(\xi,\tau),p(\xi,\tau)) \nonumber \\
			 - \left\{ [\partial_\xi q(\xi,\tau)][\partial_\xi p(\xi,\tau)] -q(\xi,\tau)[1-q(\xi,\tau)][\partial_\xi p(\xi,\tau)]^2  \right\} 
		\label{eq:sp_stoch_PI_ScontH}, \\
	\fl
	h_0(q,p) := (e^p -1)W_{+}(q) + (e^{-p}-1)W_{-}(q) \label{eq:sp_stoch_PI_ScontH0},
\end{eqnarray}
\red{where the prime in the path integral indicates the restriction to paths that satisfy $q(\cdot, \tau) \neq q_1$ ($\forall \tau \in [0,T]$). }


\subsubsection{Semiclassical approximation}
\label{subsubsec:sp_stoch_SC}

The semiclasslcal approximation can be applied in almost the same manner as was performed in section \ref{subsubsec:wm_stoch_SC} for the well-mixed model. 
We consider a stationary solution of the action represented by (\ref{eq:sp_stoch_PI_ScontS}). 
The stationary condition $\delta S = 0$ leads to partial differential equations
\begin{eqnarray}
	\fl
	\partial_\tau q &=& e^{p}W_{+}(q) - e^{-p}W_{-}(q) 
			+ \left[ \partial^{2}_{\xi}q - 2 q(1-q)\partial^{2}_{\xi} p
			-  2(1-2q)(\partial_\xi p)(\partial_\xi q) \right] \label{eq:sp_stoch_SC_eqmq}, \\
	\fl
	\partial_\tau p &=& -(e^{p}-1)W'_{+}(q) - (e^{-p}-1)W'_{-}(q)
		- \left[ \partial^{2}_{\xi} p + (1-2q)(\partial_\xi p)^2  \right]. \label{eq:sp_stoch_SC_eqmp}
\end{eqnarray}
subjected to a boundary condition $q(\cdot, 0) = q(\cdot, T) = q_3$. 
Note that when $p=0$, the equation for $q$ coincides with (\ref{eq:sp_det_eq_PDE}), 
i.e. the equation for the expectation value. 

\begin{figure}[tb]
	\centering
	\includegraphics[height=6cm]{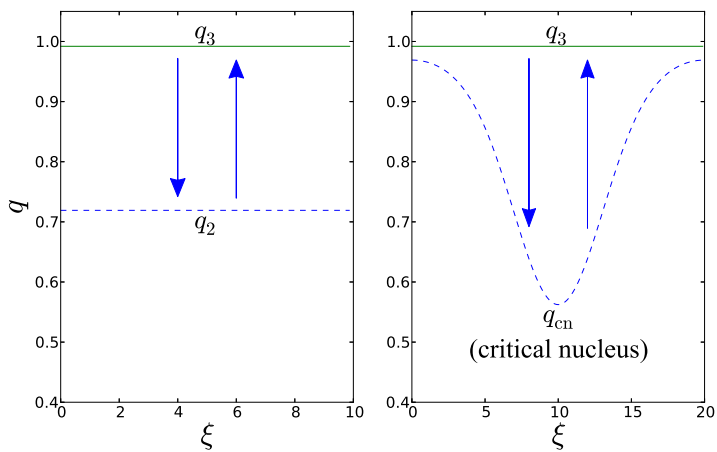}
	\caption{Schematic representation of bounce solutions. 
		\red{Left: a uniform bounce solution $\alpha$. The system changes from $q_3$ to $q_2$ and returns back to $q_3$ uniformly in space. 
		Right: a nonuniform bounce solution $\beta$. The system changes from $q_3$ to $q_{\rm cn}$ and returns back to $q_3$. }}
	\label{fig:BounceSchmtc}
\end{figure}

We now calculate $P_{q_3, {\rm sp}}(T)$ ($T \gg 1$) by considering solutions to 
the equations of motion (\ref{eq:sp_stoch_SC_eqmq}) and (\ref{eq:sp_stoch_SC_eqmp}), 
under the boundary condition $q(\xi, 0) = q(\cdot, T) = q_3$, 
and the restriction that $q$ never visits $q_1$. 
Unlike the well-mixed model, it is not easy to obtain such solutions in the spatial model.
However, we can infer qualitative features of these solutions from the analysis of the well-mixed model, 
where the solution is composed of two trajectories: a trajectory moving from $q_3$ to $q_2$ in the $p<0$ region 
and a trajectory moving from $q_2$ to $q_3$ in the $p=0$ region (see section \ref{subsubsec:wm_stoch_SC}, figure \ref{fig:scpp} right panel). 
Note that the unstable fixed point $q_2$ plays the role of a ``watershed'', i.e. a dividing point between the two stable states $q_1$ and $q_3$. 
In the spatial model, there are two kinds of unstable solutions: (a) the uniform solution $q(\xi) \equiv q_2$ and (b) the critical nucleus. 
Therefore, it is inferred that there are two kinds of bounce solutions: 
\begin{description}
	\item[(a)]
		a solution which is initially $q(\xi) \equiv q_3$, then changes into $q(\xi) \equiv q_2$, and finally returns to $q(\xi) \equiv q_3$ (see figure \ref{fig:BounceSchmtc} left panel), 
		and 
	\item[(b)]
		a solution which is initially $q(\xi) \equiv q_3$, then changes into $q = q_{\rm cn}$, and finally returns to $q(\xi) \equiv q_3$ (see figure \ref{fig:BounceSchmtc} right panel). 
\end{description}
We call the former a ``uniform bounce solution'' (or solution $\alpha$) and the latter a ``non-uniform bounce solution'' (or solution $\beta$). 
It is clear that the nonuniform bounce solution exists only when $L> L_c$. 

Let $S_{\alpha, q_3}$ and $S_{\beta, q_3}$ be the action of the bounce solutions $\alpha$ and $\beta$ calculated from (\ref{eq:sp_stoch_PI_ScontS}) - (\ref{eq:sp_stoch_PI_ScontH0}). 
If $L< L_c$, $\alpha$ is the only possible bounce solution. Hence, $P_{q_3, {\rm sp}}(T)$ is calculated to be
\begin{equation}
	P_{q_3, {\rm sp}}(T) \simeq \exp \left( -K_{\alpha, q_3} e^{-N_p\sqrt{\frac{\sigma}{\lambda}}S_{\alpha, q_3} } \ T  \right), 
\end{equation}
where $K_{\alpha,q_3}$ is a prefactor determined by the Gaussian integrals around the stationary solution $\alpha$. 
Then, $\tau_{q_3, {\rm sp}}$, the lifetime of the metastable state $q_3$ can be expressed as 
$\tau_{q_3, {\rm sp}} \simeq K_{\alpha, q_3}^{-1} \exp\left( N_p \sqrt{\sigma/\lambda} S_{\alpha, q_3} \right)$.
If $L>L_c$, there are bounce solutions $\alpha$ and $\beta$. 
By summing up all the contributions from these two bounce solutions, we obtain 
\begin{eqnarray}
	P_{q_3, {\rm sp}}(T)
	&\simeq& \sum_{n=0}^{\infty} \frac{1}{n!} \sum_{k=0}^{n} {}_{n}C_k \int_{0}^{T}dt_1 \cdots \int_{0}^{T} dt_n \nonumber \\
		&{}& \hspace{1cm}
		\times \left( -K_{\alpha, q_3} e^{-N_p \sqrt{\frac{\sigma}{\lambda}}S_{\alpha, q_3}} \right)^k \left( -K_{\beta, q_3} e^{-N_p \sqrt{\frac{\sigma}{\lambda}}S_{\beta, q_3}} \right)^{n-k}  \\
	&=& \exp\left[ -\left( K_{\alpha, q_3} e^{-N_p \sqrt{\frac{\sigma}{\lambda}} S_{\alpha, q_3}} + K_{\beta, q_3} e^{-N_p \sqrt{\frac{\sigma}{\lambda}}S_{\beta, q_3}} \right) T \right], 
\end{eqnarray}
where $K_{\alpha, q_3}$ and $K_{\beta, q_3}$ are prefactors determined by the Gaussian integrals around $\alpha$ and $\beta$ respectively. 
Then, 
\begin{equation}
	\frac{1}{\tau_{q_3, {\rm sp}}} \simeq K_{\alpha, q_3} e^{-N_p \sqrt{\frac{\sigma}{\lambda}} S_{\alpha, q_3} } + K_{\beta, q_3} e^{-N_p \sqrt{\frac{\sigma}{\lambda}}S_{\beta, q_3}}.
\end{equation}
holds. 
Under the condition $N_p\sqrt{\sigma/\lambda} \gg 1$ assumed in this paper, 
the \red{lifetime} is well-approximated by one of the two terms, which has the smaller value of the action. 
Let $\gamma$ be either $\alpha$ or $\beta$, which corresponds to the smaller value of action 
(e.g. if $S_{\alpha, q_3} > S_{\beta, q_3}$, then $\gamma = \beta$. In fact, for the case treated in the next section, we obtain $\gamma = \beta$). 
Then, we obtain $\tau_{q_3, {\rm sp}} \simeq K_{\gamma, q_3}^{-1} \exp \left( N_p \sqrt{\sigma/\lambda} S_{\gamma, q_3} \right)$.  
Thus, the expression for the lifetime is summarized as
\begin{equation}
	\tau_{q_3, {\rm sp}} \simeq 
	\cases{ 
		\frac{1}{K_{\alpha, q_3}} \exp\left( N_p \sqrt{\frac{\sigma}{\lambda}} S_{\alpha, q_3} \right) & (for $L < L_c)$ \\
		\frac{1}{K_{\gamma, q_3}} \exp\left( N_p \sqrt{\frac{\sigma}{\lambda}} S_{\gamma, q_3} \right)  & (for $L > L_c)$ \\}.
\end{equation}


\subsection{Results}
\label{subsec:sp_result}

To discuss the $L$ dependence of the lifetimes, 
we numerically calculated the action of the bounce solutions (see \ref{sec:numcalc} for the detail). 
If two kinds of bounce solutions exist (i.e. if $L > L_c$), we choose the one with the smaller action. 
The obtained action is denoted by $S_{q_i, {\rm sp}}$ ($i \in \left\{ 1,3 \right\}$). 

\begin{figure}[tb]
	\centering
	\includegraphics[width=15cm]{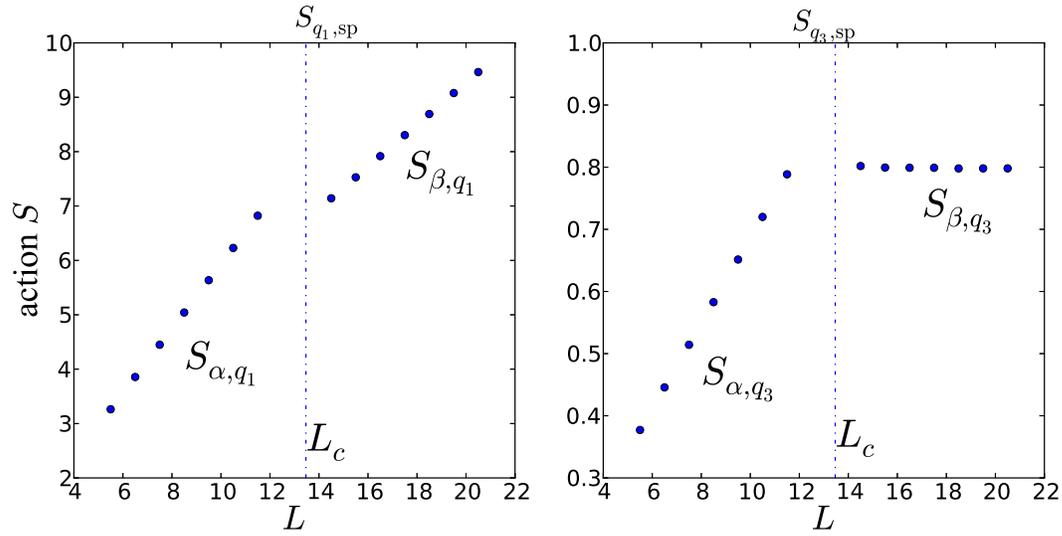}
	\caption{$L$ dependence of the action. 
		Left: $S_{q_1, {\rm sp}}$, action of the bounce solutions that determine the lifetime of $q_1$. 
		Right: $S_{q_3, \rm{sp}}$, action of the bounce solutions that determine the lifetime of $q_3$. 
		The dashed line denotes $L_c$. 
		\red{$S_{q_1, \rm{sp}}$ keeps increasing with $L$, whereas $S_{q_3, \rm{sp}}$ takes constant value when $L>L_c$. 
		Parameters: $a-c=0.4, d-b=1.0, w = 0.8, \mu_A = \mu_B = 0.005$. For this parameter set, $L_c \simeq 13.5$ and $V(q_1) > V(q_3)$ hold.}}
	\label{fig:ActionLdep}
\end{figure}

Figure \ref{fig:ActionLdep} shows the $L$ dependence of the action calculated in this manner for a specific parameter set, 
for which $V(q_1) > V(q_3)$ holds. 
If $L<L_c$, both $S_{q_1, {\rm sp}}(L)$ and $S_{q_3, {\rm sp}}(L)$ increase linearly with the system size $L$. 
A different feature appears when the system size exceeds the threshold length $L_c$, 
where the nonuniform bounce solutions become important; 
$S_{q_3, {\rm sp}}$ is now independent of $L$, 
whereas $S_{q_1, {\rm sp}}$ continues increasing with $L$. 
\red{Detailed discussion and intuitive meaning of these results are given in the next section.}
\section{Comparison and discussion}
\label{sec:discussion}

In this section, 
we discuss the spatial effect on the transitions between metastable states 
by comparing the result of the well-mixed model (section \ref{sec:wmmodel}) and that of the spatial model (section \ref{sec:spmodel}). 

In section \ref{subsec:disc_compare}, we discuss qualitative difference in the population size dependence of the lifetimes between these two studied models. 
In section \ref{subsec:disc_interpretation}, we present an intuition that explains the difference based on a ``nucleation'' process. 
In these two sections, we restrict ourselves to the case in which $V(q_1)  > V(q_3)$ \red{ unless otherwise specified} 
(for the opposite case, the same discussion holds by exchanging $q_1$ and $q_3$). 
In section \ref{subsec:disc_Lc}, the parameter dependence of the characteristic length scale $L_c$ is discussed in detail. 
\red{Section \ref{subsec_disc_migrationrate} is devoted to the interpretation of the spatial effect in terms of migration rate. }
We close this section with a remark on which metastable state is long-lived. 


\subsection{Spatial effect on strategy selection}
\label{subsec:disc_compare}

First, we summarize the results for the well-mixed model (section \ref{sec:wmmodel}). 
For the evolutionary game dynamics considered here (coordination games with mutations), 
there are two metastable states denoted by $q_1$ and $q_3$, where $q \in [0,1]$ is the proportion of A individuals in the system; 
\red{the state $q_1$ corresponds to the population dominated by strategy B, and $q_3$ corresponds to the one dominated by strategy A.}
Although the system spends an extremely long time at $q_1$ or $q_3$, it can undergo a transition from one state to the other, via an unstable state $q_2$ 
(satisfying $q_1 < q_2 < q_3$). 
The lifetime of the metastable states $q_1$ and $q_3$ are calculated to be
\begin{equation}
	\tau_{q_i, {\rm wm}} \sim \exp(N S_{q_i, {\rm wm}}) \qquad (i \in \left\{ 1,3 \right\}),	\label{eq:disc_compare_wmlifetime}
\end{equation}
where $S_{q_1, {\rm wm}}$ and $S_{q_3, {\rm wm}}$ are the action of the bounce solution 
defined by \red{(\ref{eq:wm_Stoch_SC_Sbounce2}) and (\ref{eq:wm_Stoch_SC_Sbounce}), respectively.} 
Because $S_{q_1, {\rm wm}}$ and $S_{q_3, {\rm wm}}$ are positive and independent of $N$, 
the lifetimes $\tau_{q_i, {\rm wm}}$ increases exponentially with the population size $N$. 

Next, we summarize the results for the spatial model (section \ref{sec:spmodel}). 
The lifetimes of the metastable states $q_1$ and $q_3$ in the spatial model are evaluated to be 
\begin{equation}
	\tau_{q_i, {\rm sp}} \sim \exp\left(N_p \sqrt{\frac{\sigma}{\lambda}} S_{q_i, {\rm sp}}(L) \right) \ \ (i \in \left\{ 1,3 \right\}),
	\label{eq:disc_compare_splifetime}
\end{equation}
where $S_{q_1, {\rm sp}}$ and $S_{q_3, {\rm sp}}$ are the positive functions of the system size \red{$L = M\sqrt{\lambda/\sigma}$}, and $N_p$ is the population number per patch. 
Note that $L$ is proportional to the total population number $N_p M$, where $M$ is the number of patches. 
As discussed in section \ref{subsubsec:sp_det_steady}, the spatial model has a characteristic length scale $L_c$. 
For small systems ($L<L_c$), $S_{q_3, {\rm sp}}$ grows linearly with $L$, 
whereas for large systems ($L>L_c$), it becomes independent of $L$ (see the right panel of figure \ref{fig:ActionLdep}). 
Then, from (\ref{eq:disc_compare_splifetime}) one can conclude that the lifetime $\tau_{q_3, {\rm sp}}$ grows exponentially with $L$ for $L<L_c$, 
while it is independent of $L$ for $L>L_c$. 
In other words, while the model is effectively well-mixed when $L < L_c$, 
it behaves differently from the well-mixed model when $L > L_c$. 
We stress that this difference is quite drastic due to the large factor $N_p \sqrt{\sigma / \lambda} \ (\gg 1)$ appearing in (\ref{eq:disc_compare_splifetime}). 
On the other hand, $S_{q_1, {\rm sp}}$ has linear dependence on $L$ for both $L<L_c$ and $L>L_c$ (see the left panel of figure \ref{fig:ActionLdep}). 
This implies that there is no qualitative difference in the system size dependence of $\tau_{q_1, {\rm sp}}$ between the two regions $L<L_c$ and $L>L_c$, 
which is in clear contrast to what is observed for $\tau_{q_3, {\rm sp}}$. 

\red{ 
Note that the difference between the lifetimes of $q_1$ and $q_3$ originates from that of $V(q_1)$ and $V(q_3)$, 
where $V$ is defined in (\ref{eq:sp_det_ss_potential}). 
If $V(q_1) < V(q_3)$ (the opposite situation to the above discussion), the role of $q_1$ and $q_3$ are reversed 
so that the lifetime of $q_1$ takes constant value for $L>L_c$ while the lifetime of $q_3$ keeps growing 
(the intuitive reason for this behaviour is given in the next subsection). 
The condition $V(q_1) > V(q_3)$ (resp. $V(q_1) < V(q_3)$) can be rewritten as $A_1 > A_3$ (resp. $A_1 < A_3$), where
\begin{eqnarray}
	A_1 &:=& - \int_{q_1}^{q_2} [W_{+}(q)  - W_{-}(q)] dq > 0, \\
	A_3 &:=&  \int_{q_2}^{q_3} [W_{+}(q)  - W_{-}(q)] dq > 0.
\end{eqnarray}
Note that $A_1$ and $A_3$ correspond to the two areas enclosed by the curve $W_{+}(q) - W_{-}(q)$ and the horizontal axis in figure \ref{fig:epp}. 
Without mutation ($\mu_A = \mu_B = 0$), the condition $A_1 > A_3$ (resp. $A_1 < A_3$) is clearly related to 
the condition  $q_2 > 1/2$ (resp. $q_2 < 1/2$), where $q_2 = (d-b)/(a-c+d-b)$, 
because $W_{+}(q) - W_{-}(q)$ is a cubic function of $q$. 
Thus, the condition $V(q_1) > V(q_3)$ (resp. $V(q_1) < V(q_3)$) is equivalent to $d-b > a-c$ (resp. $d-b < a-c$). 
With small mutation, although the precise condition is slightly modified, 
the above discussion still holds approximately. 
}

\red{
The observation above shows that (in the limit of small mutation ), 
the lifetimes of the metastable states are simply characterized by the relative size of $a-c$ and $d-b$ in the original game (\ref{eq:wm_model_payoffmat}). 
The value $a-c \ (> 0)$ measures the coordination advantage of playing A over playing B when matched with an A individual.  
Similarly, the value $d - b \ (> 0)$ measures the coordination advantage of playing B over playing A when matched with a B individual. 
The strategy with the larger (resp. smaller) coordination advantage is called a ``risk-dominant'' (resp. ``risk-dominated'') strategy in game theory 
\cite{NowakBook} (e.g. if $d-b > a-c$, strategy B is risk-dominant and strategy A is risk-dominated).
Therefore, our results are summarized as follows: 
the lifetime of the metastable state of the risk-dominant strategy keeps growing exponentially with the system size, 
while that of the risk-dominated one saturates after the system size hits a certain threshold. 
Hence (when the system is sufficiently large) spatial structure drastically accelerates the selection for the strategy with a larger coordination advantage
($=$risk-dominant strategy), facilitates its invasion against a risk-dominated strategy, 
and gives a prediction of how natural selection resolves coordination problems in biology. 
}

\red{
Similar spatial effect has already been found by Ellison for an evolutionary game model (Theorem 3 in \cite{Ellison1993}). 
However, direct comparison with his result is not appropriate because his model assumes one individual per each patch, 
while our model assumes $N_p (\gg 1)$ individuals per patch. 
}


\subsection{\red{The origin of the spatial effect: an intuition}}
\label{subsec:disc_interpretation}

\begin{figure}[tb]
	\begin{minipage}{0.2\hsize}
		\centering
		\includegraphics[height=4.0cm]{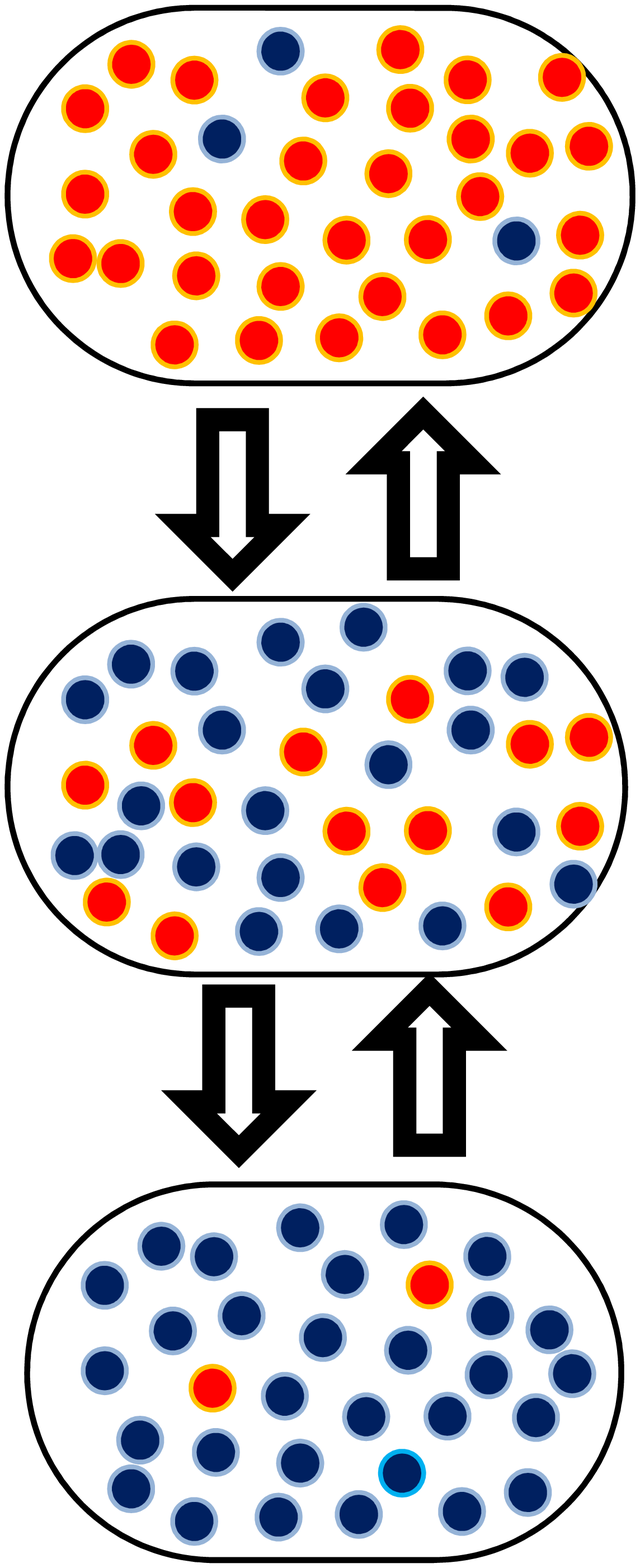}
		(a)$L< L_c$
	\end{minipage}
	\begin{minipage}{0.4\hsize}
		\centering
		\includegraphics[angle=-90,totalheight=4.0cm]{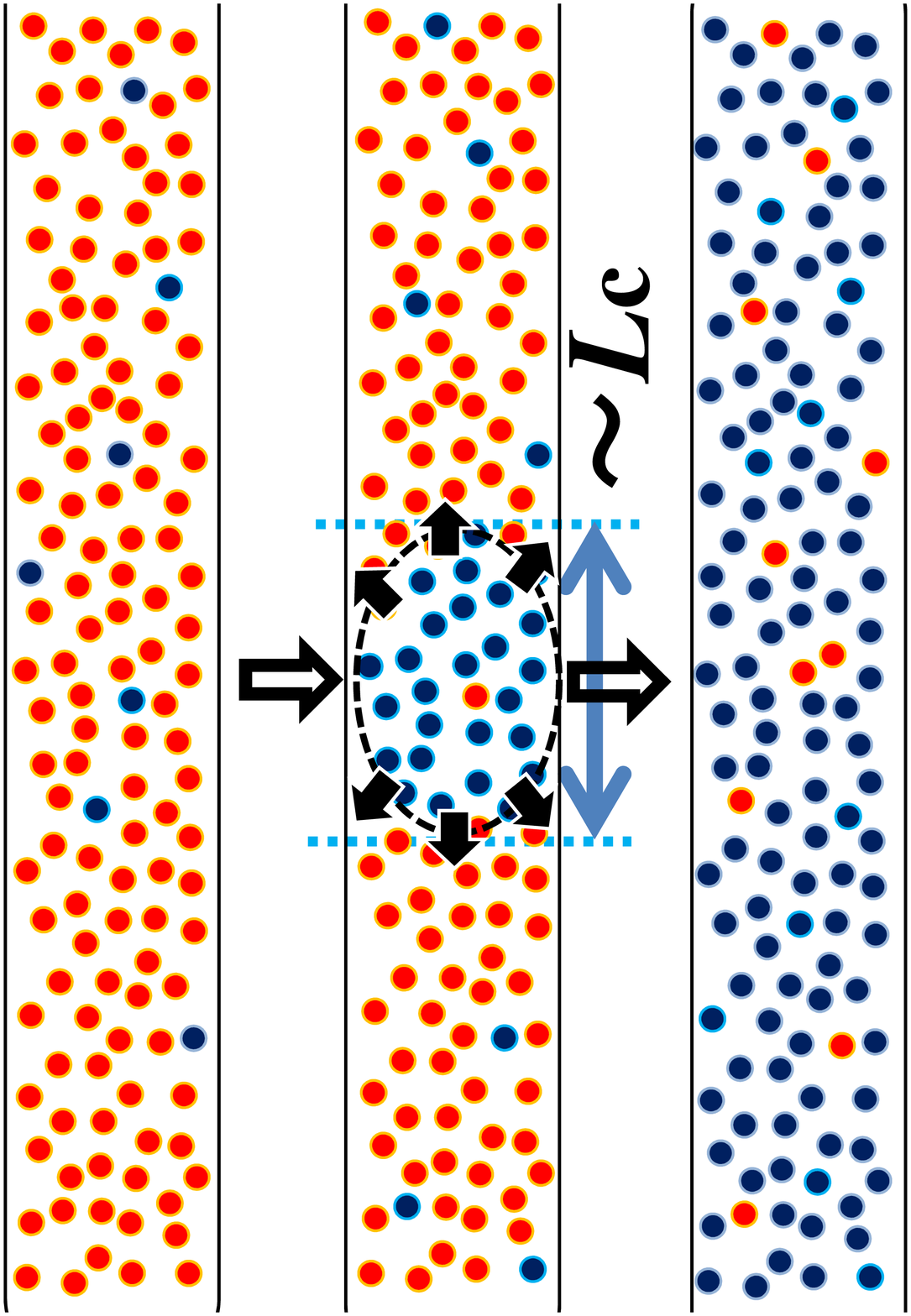}
		(b)The transition from $q_3$ to $q_1$ when $L> L_c$
	\end{minipage}
	\begin{minipage}{0.4\hsize}
		\centering
		\includegraphics[angle=-90,totalheight=4.0cm]{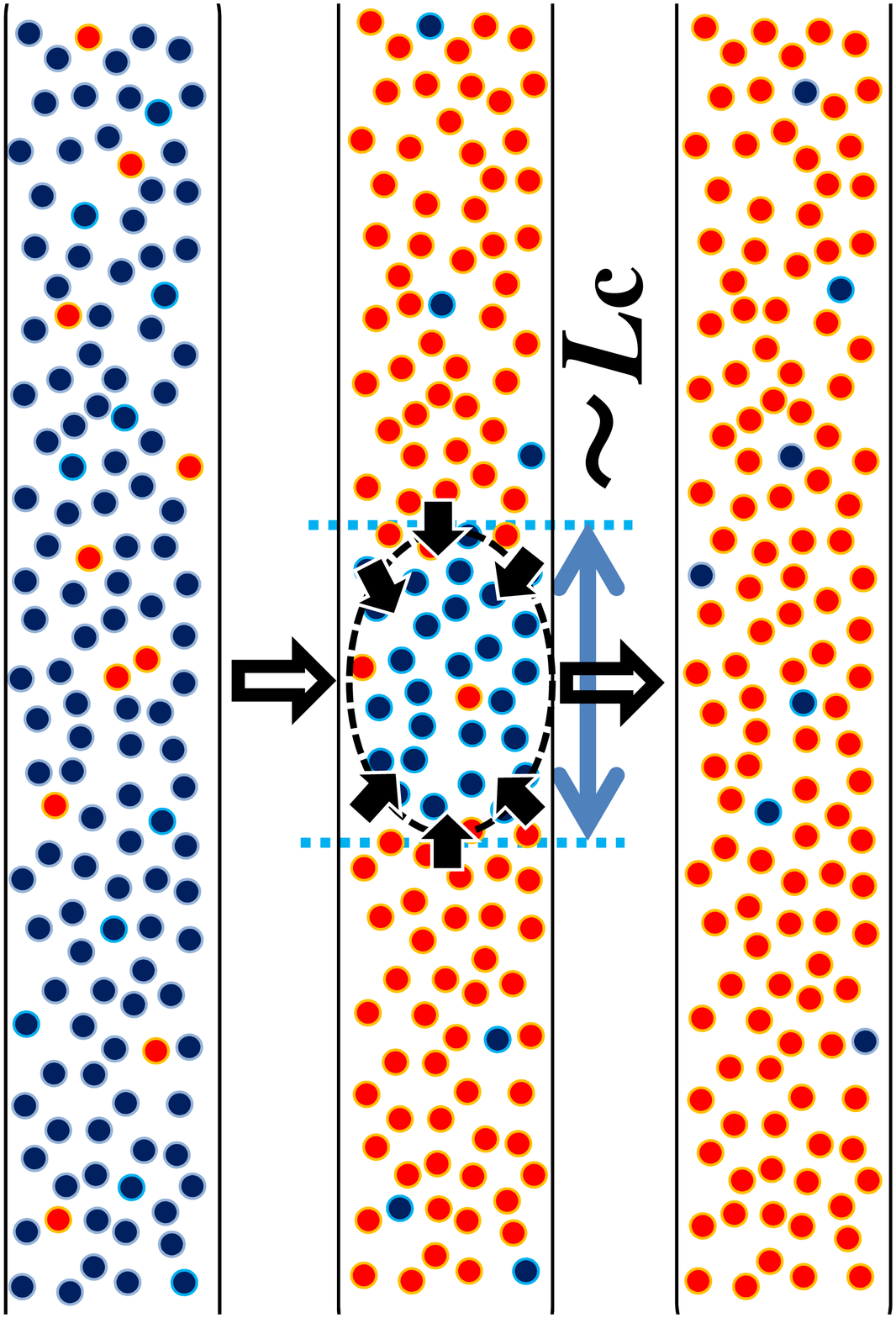}
		(c)The transition from $q_1$ to $q_3$ when $L> L_c$
	\end{minipage}
	\caption{Schematic figures illustrating the transition between metastable states in the spatial model when $V(q_1) > V(q_3)$. 
		\red{Red and blue circles represent $A$ and $B$ individuals, respectively. 
		(a) In the well-mixed model or the spatial model for $L<L_c$, transition between metastable states requires the whole system to change (almost uniformly) to cross the dividing point $q_2$. 
		In the spatial model for $L>L_c$, the transition can be triggered by crossing critical nucleus: 
		(b) For transition from $q_3$ to $q_1$ to occur, only a part of the system, whose length scale is approximately $L_c$, has to change. 
		(c) For transition from $q_1$ to $q_3$ to occur, almost the whole system has to change.  }
		}
	\label{fig:transition_schmtc}
\end{figure}

\red{Here we intuitively discuss the origin} of the qualitative differences in the system size dependence of the lifetimes 
between the well-mixed model and the spatial model. 
We first point out that during the transition from one metastable state to the other, 
the system must cross a ``dividing point'', 
which is a marginal unstable steady solution of the system located between two metastable states. 
Once the system has crossed this dividing point, it evolves with very high probability along ``deterministic'' path 
(the trajectory of the expectation value satisfying (\ref{eq:wm_det_ODE}) or (\ref{eq:sp_det_eq_PDE})), 
which quickly carries the system to the other metastable state. 
This view implies that the lifetime of a metastable state is mostly determined by the difficulty of crossing a dividing point from a given metastable state. 
Then, the system size dependence of the lifetimes can be intuitively explained as follows (again, we restrict our consideration to the case $V(q_1) > V(q_3)$): 
\begin{itemize}
	\item 
		In the well-mixed model or the spatial model for $L< L_c$, the unstable state $q_2$ plays the role of a dividing point. 
		To make a transition from one metastable state to the other, 
		the entire system has to cross the dividing point $q_2$ (figure \ref{fig:transition_schmtc}(a)). 
		As the system size increases, the probability that the system changes from a given metastable state to the dividing point 
		decays exponentially with $N$ (or $L$), resulting in the exponential dependence of the lifetimes on $N$ (or $L$). 
	\item 
		In the spatial model for $L > L_c$, the critical nucleus also plays the role of a dividing point. 
		This indicates that the transition from one metastable state to the other can be caused by ``nucleation'', i.e. transition via the critical nucleus. 
		The details of the nucleation process depend on the direction of transition. 
		\begin{itemize}
			\item 
				For a transition from $q_3$ to $q_1$ to occur, it is sufficient for the system to change from $q_3$ to the critical nucleus, 
				which means that the change of only a part of the system, whose length scale is approximately $L_c$, is sufficient 
				(\red{see figure \ref{fig:transition_schmtc} (b) and the spatial form of the critical nucleus in the right panel of figure \ref{fig:cnuc} }). 
				The difficulty of this change does not depend on $L$, and hence neither does the difficulty of transition from $q_3$ to $q_1$. 
				Therefore, the lifetime of $q_3$ does not grow with $L$. 
			\item 
				For a transition from $q_1$ to $q_3$ to occur, it is necessary for the system to change from $q_1$ to the critical nucleus, 
				which means that almost the entire system has to change (figure \ref{fig:transition_schmtc}(c)). 
				This change becomes more difficult as $L$ grows, and  so does the transition from $q_1$ to $q_3$. 
				Hence, the lifetime of $q_1$ grows exponentially with $L$. 
		\end{itemize}
\end{itemize}
It should be noted that the existence of two kinds of transition paths (shown in figure \ref{fig:transition_schmtc}) is directly related to the two kinds of bounce solutions, 
i.e. the uniform bounce solution and the non-uniform bounce solution discussed in section \ref{subsubsec:sp_stoch_SC}. 
We stress that the essential change of the transition process is described by the change in the bounce solutions.

\red{Note that when $V(q_1) < V(q_3)$, the behaviour of the lifetime of $q_1$ and $q_3$ is reversed. 
This is because the form of critical nucleus becomes ``upside down'' compared with right panel of figure \ref{fig:cnuc}, 
so that the role of strategy A (red) and B (blue) in figure \ref{fig:transition_schmtc} should be replaced by each other. 
}

\begin{figure}
	\centering
	\includegraphics[width=14cm]{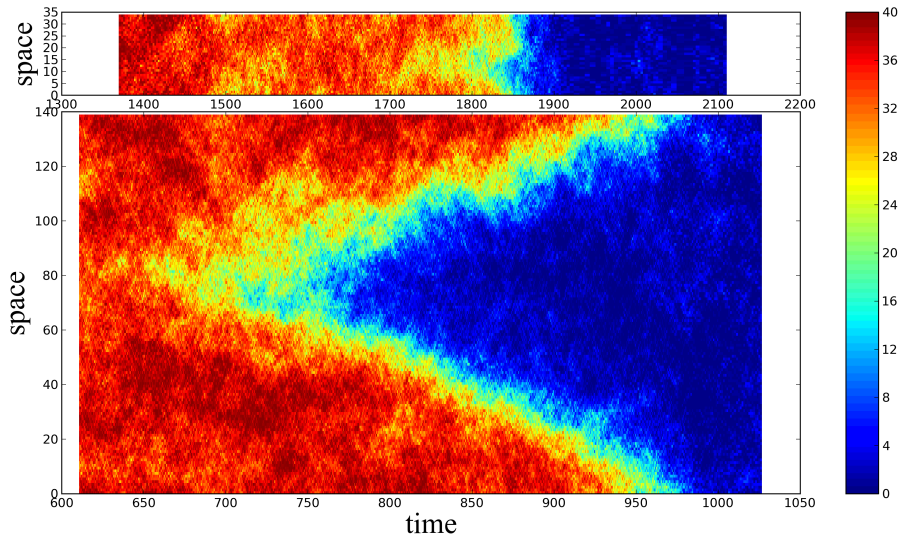}	
	\caption{Results of Monte Carlo simulation \red{of master equation (\ref{eq:sp_model_mastereq})}. Top: $M = 35$, Bottom: $M = 140$. 
		\red{ The color shows the number of $A$ individuals in each patch ($N_p = 40$). 
		Parameters: $N_p = 40, a-c = 0.4, d-b = 1.0, w = 0.1, \mu_A = \mu_B = 0.005, 
		\lambda = N_p, \sigma = 2N_p$ (for a detailed description of the parameters, see section \ref{subsec:wm_model} and \ref{subsec:sp_model}).
		For this parameter set, $N_p q_1 \simeq 1.01, N_p q_3 \simeq 36.8$ and $L_c \simeq 50.1$ (from (\ref{eq:disc_Lc_Lc})). 
		In terms of the patch number $M$, $L_c \simeq 50.1$ is equivalent to $M_c := \sqrt{\sigma/\lambda}L_c \simeq 70.8$. 
		It can be seen that for a small system (the upper figure, $M < M_c$), the transition process proceeds almost uniformly in space, 
		while in a large system (the lower figure, $M>M_c$), the transition starts from a small region and then spread across the whole system (nucleation), 
		as explained in figure \ref{fig:transition_schmtc}.} }
	\label{fig:transition_MC}
\end{figure}

To confirm the intuition described above, we performed a Monte Carlo simulation of the spatial model, 
specified by the master equation (\ref{eq:sp_model_mastereq}) employing the Gillespie algorithm \cite{Gillespie}. 
Figure \ref{fig:transition_MC} shows results of the simulation. 
\red{For a small system (the upper panel, $L<L_c$), 
the transition from one metastable state (the red region) to the other state (the blue region) occurs almost uniformly. 
For a large system (the lower panel, $L>L_c$), however, the transition proceeds via nucleation; 
transition occurs first in a small region and then it spreads over the whole system. }

We point out that the intuition presented above suggests the generality of our results; 
although we have demonstrated spatial effect on bistable evolutionary games with a specific update rule, 
the qualitative behaviour of the lifetimes is expected to be the same for models with other update rules as long as the underlying evolutionary game is bistable. 
This is because our results rely on the presence of a critical nucleus (which has a characteristic length scale $L_c$), 
which in turn originates from the structure of the phase portrait (see figures \ref{fig:epp} and \ref{fig:cnuc}). 
It can be easily shown that models with other update rules (e.g. a pairwise comparison process with other form of $p(\Delta \Pi)$ or the Moran process) 
also generate phase portraits with the same structure as figures \ref{fig:epp} and \ref{fig:cnuc}, 
provided that payoff matrix elements satisfy (\ref{eq:wm_model_coordination}) and mutation rates are sufficiently small.


\subsection{Characteristic length scale $L_c$}
\label{subsec:disc_Lc}

The characteristic length $L_c$ in the spatial model is an important quantity 
that determines the population size dependence of the lifetimes. 
In this section, we discuss its parameter dependence in detail. 
The characteristic length $L_c$ is calculated in section \ref{subsubsec:sp_det_steady}, and is given by 
\begin{equation}
	L_c = \frac{2\pi}{\sqrt{W'_{+}(q_2)- W'_{-}(q_2)}}.
	\label{eq:disc_Lc_Lc}
\end{equation}

To observe the parameter dependence more clearly, we derive an approximate expression for $L_c$: 
because $q_1, q_2$ and $q_3$ are the solutions of the cubic equation 
\red{$W_{+}(q)-W_{+}(q)=0$}, 
we obtain an alternative expression for \red{$W_{+}(q)-W_{-}(q)$ (see (\ref{eq:wm_det_cubic}))}: 
\red{
\begin{equation}
	\fl W_{+}(q)- W_{-}(q) = - w\left( 1- \frac{\mu_A + \mu_B}{2} \right) (a-b-c+d) (q-q_1)(q-q_2)(q-q_3).
\end{equation}
}
By using (\ref{eq:disc_Lc_Lc}), we obtain
\begin{equation}
	L_c = 2\pi \left/\sqrt{w \left(1-\frac{\mu_A + \mu_B}{2}\right)(a-b-c+d)(q_3-q_2)(q_2-q_1)} \right. .
\end{equation}
Assuming that $\mu_A$ and $\mu_B$ are sufficiently small, we arrive at a concise approximate form
\begin{equation}
	L_c \simeq 2\pi \sqrt{\frac{1}{w} \left( \frac{1}{a-c} + \frac{1}{d-b} \right)  }, 
	\label{eq:disc_Lc_approx}
\end{equation}
from which we can conclude that $L_c$ decreases with $w, a-c$ and $d-b$. 

\begin{figure}
	\centering
	\includegraphics[width=8cm]{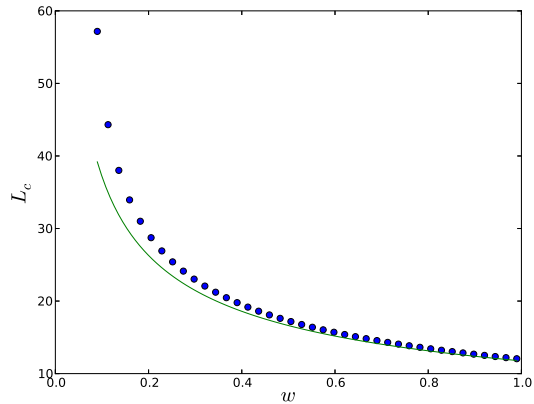}
	\caption{$w$ dependence of $L_c$ ($a-c=0.4, d-b=1.0, \mu_A = \mu_B = 0.005$).
		circle: equation (\ref{eq:disc_Lc_Lc}), solid line: approximate expression (\ref{eq:disc_Lc_approx}).}
	\label{fig:LcWdep}
\end{figure}

Figure \ref{fig:LcWdep} shows the $w$ dependence of $L_c$, 
where the circles and solid line denote the results obtained from numerical evaluation of (\ref{eq:disc_Lc_Lc}) and 
the approximate expression (\ref{eq:disc_Lc_approx}), respectively. 
One can see that the present approximation works well as $w$ increases. 

\begin{figure}
	\centering
	\includegraphics[width=15cm]{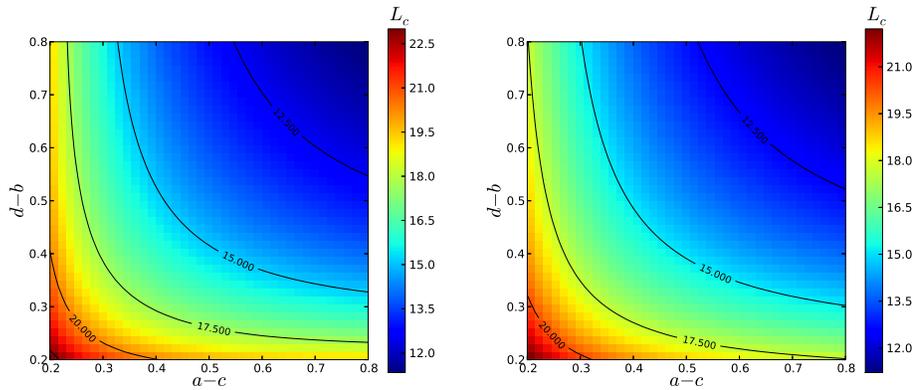}
	\caption{payoff matrix dependence of $L_c$ ($w=0.8, \mu_A = \mu_B = 0.005$)
		Left: equation (\ref{eq:disc_Lc_Lc}), Right: approximate expression (\ref{eq:disc_Lc_approx})}
	\label{fig:LcPFdep}
\end{figure}

Figure \ref{fig:LcPFdep} shows the payoff matrix dependence of $L_c$; 
the left and right panels are obtained from the numerical evaluation of (\ref{eq:disc_Lc_Lc}) 
and the approximate expression (\ref{eq:disc_Lc_approx}), respectively. 
One can see that $L_c$ decreases as $a-c$ and $d-b$ increases, 
and that the approximate expression (\ref{eq:disc_Lc_approx}) captures the qualitative feature well. 

\subsection{Migration rate dependence}
\label{subsec_disc_migrationrate}

\red{
For comparison with a real systems, it is important to judge whether the system under consideration can be regarded as well-mixed or spatial. 
In this section, we show the criterion in terms of migration rate $\sigma$, and examine the effect of $\sigma$ on the lifetimes of the metastable states. 

As was pointed out in section \ref{subsec:disc_compare}, the system is effectively well-mixed if $L<L_c$. 
Because $L$ is defined by $L := \tilde{L}/\sqrt{D} = M \sqrt{\lambda/\sigma} $ ($M:$ the number of patches),
the condition $L<L_c$ is equivalent to 
\begin{equation}
	M < \sqrt{\frac{\sigma}{\lambda}} L_c. 
\end{equation}
Note that $L_c$ depends only on game theoretical  parameters given in table \ref{tab:wm_model_param} but not on $\sigma$ or $\lambda$ (see (\ref{eq:disc_Lc_Lc})). 
Therefore, in terms of the migration rate $\sigma$, the system is effectively well-mixed if
\begin{equation}
	\sigma/ \lambda > M^2 / L_{c}^{2},  
	\label{eq:disc_gameth_sigmac}
\end{equation}
i.e. if the ratio between migration rate and update rate is larger than the value determined by $M$ and game theoretical parameters. 
On the other hand, when $\sigma$ is small so that (\ref{eq:disc_gameth_sigmac}) is not satisfied, 
the system should be described as the spatial model. 
Then, the transition between the metastable states is affected by the nucleation process described in section \ref{subsec:disc_interpretation}. 

Finally, we discuss the $\sigma$ dependence of the lifetimes of the metastable states. 
We focus on the exponents of the lifetimes
\begin{equation}
	N_p \sqrt{\frac{\sigma}{\lambda}} \ S_{q_i, {\rm sp}}\left(M \sqrt{\frac{\lambda}{\sigma}}\right)
	\label{eq:disc_gameth_LifetimeExponent}
\end{equation}
(see (\ref{eq:disc_compare_splifetime}) ). 
The lifetimes take constant values when $\sigma$ is large so that (\ref{eq:disc_gameth_sigmac}) is satisfied, 
i.e. when the system is effectively well-mixed. 
This is because $S_{q_i,{\rm sp}}(L)$ grows linearly with $L$ when $L<L_c$, and therefore
factors depending on $\sigma$ cancel out (see (\ref{eq:disc_gameth_LifetimeExponent})). 
When $\sigma$ becomes smaller than the critical value in (\ref{eq:disc_gameth_sigmac}), 
the lifetimes decrease as $\sigma$ decreases. 
Thus, it can be concluded that spatial \red{structure} accelerates transition between metastable states, 
leading to drastic reduction of the lifetimes compared with those for the well-mixed model. 
}


\subsection{Which is the long-lived metastable state?}
\label{subsec_disc_longlived}

\red{When we study coordination games, }
it is important to determine which metastable state has the longer lifetime 
because it gives a criterion for which one of A and B is better off \cite{KMR}.  
We briefly discuss the spatial effect on this problem. 

For the well-mixed model, 	the metastable state with the larger value of action $S_{q_i, {\rm wm}}$ has the longer lifetime
as can be seen from (\ref{eq:disc_compare_wmlifetime}). 
For the spatial model with a sufficient size ($L>L_c$), on the other hand, 
the metastable state with the larger value of potential $V(q_i)$ has the longer lifetime 
(see section \ref{subsec:sp_result} and \ref{subsec:disc_compare}). 

In the limit of small mutation rates ($\mu_A, \mu_B \rightarrow 0$), it can be easily shown that these two conditions coincide: 
\red{$V(q_1) > V(q_3)$ and $S_{q_1, {\rm wm}} > S_{q_3, {\rm wm}}$ are, respectively, equivalent to the condition $d-b > a-c$, 
suggesting that the risk-dominant strategy has the longer lifetime, which is consistent with previous studies \cite{KMR,Ellison1993}.} 
However, in the presence of asymmetric mutation ($\mu_A, \mu_B > 0$, $\mu_A \neq \mu_B$), the conditions are not necessarily equivalent. 
This observation indicates that long-lived metastable state may switch from one state to the other as the system size increases. 
Detailed discussion is left for a future study.

\section{Summary}
\label{sec:summary}

In this paper, we have evaluated the lifetimes of metastable states in bistable evolutionary games 
by utilizing the path integral method and the semiclassical approximation. 
It has been shown that spatial \red{structure} qualitatively changes the system size dependence of the lifetimes \red{of the metastable states}: 
For the model without spatial \red{structure} (the well-mixed model), 
we have shown that the lifetimes of the metastable states grow exponentially with the total population size $N$. 
On the other hand, for the model with spatial \red{structure} (the spatial model), 
we have shown that there exists a threshold length $L_{\rm c}$ 
across which the system size $L$($\propto$ the total population size) dependence of the lifetimes changes. 
For $L<L_{\rm c}$, the lifetimes of the two metastable states grow exponentially with the system size,  
whereas for $L>L_{\rm c}$, the lifetime of one metastable state remains constant, while that of the other continues growing exponentially with $L$. 
This significant change in the system size dependence can be intuitively explained by the \red{presence} of critical nuclei; 
for large systems ($L>L_{\rm c}$) the transition is induced by ``nucleation'' via critical nuclei. 

We stress that the present method allows semi-quantitative calculation of the lifetimes taking into account large fluctuations. 
Although we considered specific models, the present method can be easily applied to other models of the evolutionary game theory. 
Extension toward evolutionary games on higher-dimensional lattices or complex networks is an important future problem. 

\appendix

\section{Stochastic processes and path integral}
\label{sec:SPandPI}

In this appendix, we will derive a path integral expression for the transition probability 
used in section \ref{subsubsec:wm_stoch_PI} and \ref{subsubsec:sp_stoch_PI}. 
The derivation is based on \cite{LefevreBiroli}. 

We consider a general continuous time Markov process on the discrete states $n\in\mathbb{N}$. 
The rates at which processes $n \rightarrow n+ r \ (r \in R \subset \mathbb{Z} - \left\{ 0 \right\})$ occur are given by $W_r(n)$. 
By setting \red{$W_{\pm 1}(n)=W_{\pm}(n/N)$} and $W_r(n) = 0$ (for $|r| \geq 2$), we obtain the well-mixed model discussed in section \ref{sec:wmmodel}. 
Let $t \in [0,T]$ be time. 
Our aim is to calculate $p(n_{\rm fin}, T | n_{\rm ini}, 0) $, the probability that $n= n_{\rm fin}$ at $t = T$ given that $n = n_{\rm ini}$ at $t = 0$ 
(other conditions can be added, such as restriction of the path). 
 
First, we discretize the time interval $[0,T]$ into $K (\gg 1)$ small fractions: 
\begin{equation}
	\Delta t := T/K, \qquad t_j := j \Delta t,  \qquad n_j  := n(t_j) \qquad (j \in \left\{ 0,1, \cdots, K \right\}), 
\end{equation}
where $n_0 := n_{\rm ini}$ and $n_{K} := n_{\rm fin}$. 
Let $P(\Delta n | n)$ be the probability that the system changes from $n$ to $n + \Delta n$ in one discrete time step. 
Then, $p(n_{\rm fin}, T | n_{\rm ini}, 0) $ can be approximated by
\begin{eqnarray}
	\fl
	&{} p(n_{\rm fin}, T | n_{\rm ini}, 0)  \nonumber \\
	\fl
	&\simeq \sum_{n_1, \cdots, n_{K-1} = 0}^{\infty} 	P(n_1 - n_0|n_0) \cdots P(n_K - n_{K-1}|n_{K-1}) \nonumber \\
	\fl
	&= \sum_{n_1, \cdots, n_{K-1} = 0}^{\infty} \sum_{\Delta n_0, \cdots, \Delta n_{K-1}} \prod_{j=0}^{K-1} \delta_{n_{j+1}-n_j, \Delta n_j} P(\Delta n_j | n_j) \nonumber \\
	\fl
	&= \sum_{\Delta n_0, \cdots, \Delta n_{K-1}} \int dn_1 \cdots dn_{K-1} \prod_{j=0}^{K-1}
		\delta(n_{j+1}-n_j-\Delta n_j)P(\Delta n_j|n_j).
\end{eqnarray}

By substituting 
\begin{equation}
	\fl
	\delta(n_{j+1}-n_j - \Delta n_j) = \int_{-i\infty}^{i\infty} \frac{dp_{j+1}}{2\pi i } e^{p_{j+1}(\Delta n_j - n_{j+1} + n_j)} \ \ (j \in \left\{0,1,\cdots, K-1 \right\}),
\end{equation}
we obtain 
\begin{eqnarray}
	\fl
	p(n_{\rm fin}, T | n_{\rm ini}, 0) &\simeq \int dn_1 \cdots dn_{K-1} \int \frac{dp_1}{2\pi i} \cdots \frac{dp_K}{2\pi i} \nonumber \\
		& \times \exp\left[-\sum_{j=0}^{K-1} p_{j+1}(n_{j+1} - n_j)\right] \prod_{j=0}^{K-1} \sum_{\Delta n_j } e^{p_{j+1} \Delta n_j} P(\Delta n_j | n_j).
\end{eqnarray}
Since
\begin{eqnarray}
	\sum_{\Delta n_j} e^{p_{j+1} \Delta n_j} P(\Delta n_j | n_j)  &= \sum_{r \in R} e^{r p_{j+1}} W_r(n_j)\Delta t + \left( 1 - \sum_{r \in R} W_r(n_j) \Delta t \right) \nonumber \\
	&\simeq \exp\left[ \sum_{r \in R} \left(e^{r p_{j+1}}  -1 \right) W_r(n_j)\Delta t \right],
\end{eqnarray}
holds for sufficiently small time interval $\Delta t$, 
we arrive at 
\begin{eqnarray}
	\fl
	p(n_{\rm fin}, T | n_{\rm ini}, 0) \simeq \nonumber \\
	\fl
	\int \prod_{i=1}^{K-1} dn_i \int \prod_{j=1}^{K} \frac{dp_j}{2\pi i} 
		\exp \left\{ -\sum_{k=0}^{K-1} \left[ p_{k+1}(n_{k+1}-n_k) -\Delta t \sum_{r\in R}(e^{rp_{k+1}}-1)W_r(n_k) \right]  \right\}.
\end{eqnarray}
Taking the limit $K \rightarrow \infty$ yields 
\begin{eqnarray}
	p(n_{\rm fin}, T | n_{\rm ini}, 0)
		&=& \int_{n(0) = n_{\rm ini}, n(T) = n_{\rm fin}} \mathcal{D}n \mathcal{D}p \exp(-S[n(\cdot),p(\cdot)]), \\
	S[n(\cdot),p(\cdot)] &:=& \int_{0}^{T} dt \ \left[ p(t) \partial_t n(t) - H(n(t),p(t)) \right], \\
	H(n,p) &:=& \sum_{r\in R}(e^{rp}-1)W_r(n).
\end{eqnarray}

The derivation for models with spatial degrees of freedom proceeds almost in parallel with the derivation shown above. 
\red{
We consider one-dimensional array of $M$ patches with periodic boundary condition. 
The state of the system is specified by $\boldsymbol{n} := (n^{1}, n^{2}, \cdots , n^{M})$. 
We assume that local processes occur as specified above: 
the process $\boldsymbol{n} \rightarrow \boldsymbol{n}+ r\boldsymbol{e}_i \ (r \in R \subset \mathbb{Z} - \left\{ 0 \right\})$ occurs at rate $W_r(n^{i})$, 
where 
\begin{equation}
	\boldsymbol{e}_i := (0, \cdots, 0, \stackrel{i}{\breve{1}}, 0, \cdots, 0). 
\end{equation}
In addition to the local processes, we assume there are migration processes between neighbouring patches: 
for $|i-j| = 1$, the process $\boldsymbol{n} \rightarrow \boldsymbol{n} - \boldsymbol{e}_i + \boldsymbol{e}_j$ 
occurs at rate $W_{\rm m}(n^{i}, n^{j})$. 
By replicating the discussion above, we obtain
\begin{eqnarray}
	\fl
	p(\boldsymbol{n}_{\rm fin}, T | \boldsymbol{n}_{\rm ini}, 0) 
		&\simeq \int d\boldsymbol{n}_1 \cdots d\boldsymbol{n}_{K-1} \int \frac{d\boldsymbol{p}_1}{2\pi i} \cdots \frac{d\boldsymbol{p}_K}{2\pi i} \nonumber \\
	& \times \exp\left[-\sum_{j=0}^{K-1} \boldsymbol{p}_{j+1}(\boldsymbol{n}_{j+1} - \boldsymbol{n}_j)\right] \prod_{j=0}^{K-1} 
				\sum_{\Delta \boldsymbol{n}_j } e^{\boldsymbol{p}_{j+1} \Delta \boldsymbol{n}_j} P(\Delta \boldsymbol{n}_j | \boldsymbol{n}_j).
\end{eqnarray}
Since
\begin{eqnarray}
	\fl  e^{\boldsymbol{p}_{k+1} \Delta \boldsymbol{n}_k} P(\Delta \boldsymbol{n}_k | \boldsymbol{n}_k) \nonumber \\
	\fl = \Delta t \sum_{i=1}^{M} \sum_{r\in R} e^{r p_{k+1}^{i}} W_r(n_{k}^{i}) 
			+ \Delta t \sum_{i,j=1, |i-j|=1}^{M} e^{-p_{k+1}^{i}+ p_{k+1}^{j}} W_{\rm m}(n_{k}^{i}, n_{k}^{j})  \nonumber \\
	\fl	\qquad + \left( 1 - \Delta t \sum_{i=1}^{M} \sum_{r\in R} W_r(n_{k}^{i}) - \Delta t \sum_{i,j=1, |i-j|=1}^{M} W_{\rm m}(n_{k}^{i}, n_{k}^{j})  \right) \nonumber \\
	\fl \simeq \exp\left\{ \Delta t \sum_{i =1}^{M} \sum_{r\in R}(e^{p_{k+1}^{i}}-1)W_r(n_{k}^{i}) 
		+ \Delta t \sum_{i,j=1, |i-j|=1}^{M} (e^{-p_{k+1}^{i}+ p_{k+1}^{j}} - 1) W_{\rm m}(n_{k}^{i}, n_{k}^{j})  \right\}
\end{eqnarray}
holds for sufficiently small $\Delta t$, we arrive at 
\begin{eqnarray}
	\fl
	p(\boldsymbol{n}_{\rm fin}, T | \boldsymbol{n}_{\rm ini}, 0) \simeq 
		\int d\boldsymbol{n}_1 \cdots d\boldsymbol{n}_{K-1} \int \frac{d\boldsymbol{p}_1}{2\pi i} \cdots \frac{d\boldsymbol{p}_K}{2\pi i} \nonumber \\
	\fl	 \qquad \times \exp\left\{-\sum_{k=0}^{K-1} \left[ \boldsymbol{p}_{k+1}(\boldsymbol{n}_{k+1} - \boldsymbol{n}_k) - H(\boldsymbol{n}_k, \boldsymbol{p_{k+1}})\Delta t \right] \right\},
\end{eqnarray}
where
\begin{equation}
	\fl H(\boldsymbol{n}, \boldsymbol{p}) := 
	\sum_{i =1}^{M} \sum_{r\in R}(e^{p^{i}}-1)W_r(n^{i})  +  \sum_{i,j=1, |i-j|=1}^{M} (e^{-p^{i}+ p^{j}} - 1) W_{\rm m}(n^{i}, n^{j}). 
\end{equation}

}
\red{
\section{Stability of non-uniform solutions}
\label{sec:Stability_cn}

In this appendix, we outline the proof that a non-uniform steady solution for (\ref{eq:sp_det_eq_PDE}) is unstable. 
Let $q_{\rm cn}(\cdot)$ be a non-uniform solution for (\ref{eq:sp_det_eq_PDE}). 
Linearization of (\ref{eq:sp_det_eq_PDE}) around $q_{\rm cn}$ by substituting $q(\xi,\tau) = q_{\rm cn}(\xi) + \delta q(\xi,\tau)$ yields
\begin{equation}
	\partial_{\tau} \delta q(\xi,\tau) = \mathcal{L} \delta q(\xi,\tau), \qquad \mathcal{L} := \partial_{\xi}^2 + W'_{+}(q_{\rm cn}(\xi)) - W'_{-}(q_{\rm cn}(\xi)). 
\end{equation}
If all the eigenvalues of $\mathcal{L}$ are negative, $q_{\rm cn}$ is linearly stable, and otherwise $q_{\rm cn}$ is linearly unstable. 
It can be easily shown that $\delta q_0 := dq_{\rm cn}/d\xi$ is an eigenfunction with zero eigenvalue (called a zero mode) of $\mathcal{L}$. 
Because $\delta q_0$ has two nodes, one can show (with the help of the theory of periodic Sturm-Liouville problems \cite{CoddingtonLevinson}) 
that there must be one nodeless eigenmode with a positive eigenvalue. 
Hence, $q_{\rm cn}$ is linearly unstable.

The above result indicates that a non-uniform steady solution represents a marginal state located at the boundary between the metastable states $q_1$ and $q_3$. 
If a positive perturbation $\delta q(\xi) ( > 0 \mbox{ for all } \xi)$ is added to $q_{\rm cn}$, the system is driven toward $q_3$ following the deterministic equation (\ref{eq:sp_det_eq_PDE}). 
On the other hand, if a negative perturbation $\delta q(\xi) ( < 0 \mbox{ for all } \xi)$ is added, the system is driven toward $q_1$. 

}
\section{Numerical calculation}
\label{sec:numcalc}

\red{
In this appendix, we briefly describe the numerical method used to obtain the action of the bounce solution in the spatial model (section \ref{subsubsec:sp_stoch_SC}). 
As in section \ref{subsubsec:sp_stoch_SC}, we calculate bounce solutions that determines the lifetime of $q_3$ (almost the same discussion applies to $q_1$) .

First, note that for both the solutions $\alpha$ and $\beta$, it is not necessary to consider the second half of the transition 
(transition from $q_2$ to $q_3$ for the bounce solution $\alpha$, and transition from $q_{\rm cn}$ to $q_3$ for the bounce solution $\beta$) 
since $p \equiv 0$ for these trajectories and hence they do not contribute to the action (see (\ref{eq:sp_stoch_PI_ScontS})-(\ref{eq:sp_stoch_PI_ScontH0})). 
This is because they corresponds to trajectories obeying the ``deterministic'' partial differential equation (\ref{eq:sp_det_eq_PDE}), 
which can be obtained from (\ref{eq:sp_stoch_SC_eqmq}) and (\ref{eq:sp_stoch_SC_eqmp}) by setting $p \equiv 0$. 
Hence, it is sufficient to consider the first half of the transition of the the bounce solutions $\alpha$ and $\beta$, 
which we call solutions $\alpha'$ and $\beta'$ respectively, for the calculation of $S_{\alpha} (= S_{\alpha'})$ and $S_{\beta} (= S_{\beta'})$. 
These solutions are obtained by solving the partial differential equations (\ref{eq:sp_stoch_SC_eqmq}) and (\ref{eq:sp_stoch_SC_eqmp})
\begin{eqnarray}
	\partial_\tau  q &=& F(q,p) + \left[ \partial^{2}_{\xi}q - 2 q(1-q) \partial^{2}_{\xi} p	-  2(1-2q)(\partial_\xi p)(\partial_\xi q) \right] \label{eq:num_eqmq},  \\
	\partial_\tau p &=& G(q,p) - \left[ \partial^{2}_{\xi} p + (1-2q)(\partial_\xi p)^2  \right] \label{eq:num_eqmp},
\end{eqnarray}
where $F(q,p) := e^{p}W_{+}(q) - e^{-p}W_{-}(q)$ and $G(q,p) := -(e^{p}-1)W'_{+}(q) - (e^{-p}-1)W'_{-}(q)$,  
subjected to boundary conditions 
\begin{description}
	\item[$\alpha'$:] $q(\xi, 0) = q_3$ and $q(\xi,T) = q_2$ ($\forall \xi \in [0,L]$) 
	\item[$\beta'$:] $q(\xi, 0) = q_3$ and $q(\xi, T) = q_{\rm cn}(\xi)$ ($\forall \xi \in [0,L]$)
\end{description}
where $T$ is taken to be sufficiently large ($ T \gg 1$). 
This problem was solved numerically as follows. 
}

We discretize time and space as
\begin{eqnarray}
	\tau_i := i \Delta t, \qquad \Delta t := T/N_T \qquad (i \in \left\{ 0,1, \cdots, N_T -1 \right\}) \\
	\xi_j := j \Delta x, \qquad \Delta x := L/N_L \qquad (j \in \left\{ 0,1, \cdots, N_L-1 \right\}) \\
	q(\xi_j, \tau_i) \leftrightarrow q_{i,j}, \qquad p(\xi_j, \tau_i) \leftrightarrow p_{i,j}, 
\end{eqnarray}
where $N_T$ and $N_L$ are the division numbers of the time and spatial coordinates, respectively. 
Note that $q_{0,j}, \ q_{N_T-1,j} \ (j \in \left\{ 0,1, \cdots, N_L-1 \right\})$ are given as the boundary condition. 
We then discretize (\ref{eq:num_eqmq}) and (\ref{eq:num_eqmp}) into 
the following difference equations: 
For $i \in \left\{ 1,2, \cdots, N_T -2 \right\} , \ j \in \left\{ 0,1, \cdots, N_L-1 \right\}$
\begin{eqnarray}
	\fl
	\frac{q_{i+1,j } - q_{i-1,j}  }{2\Delta t}  &=& F(q_{i,j}, p_{i,j}) \nonumber \\
	\fl
			&+& \left[ \frac{q_{i,j+1} - 2 q_{i,j} + q_{i,j-1}}{(\Delta x)^2}   
						-2 (1-2 q_{i,j})\frac{q_{i,j+1}-q_{i,j-1}}{2\Delta x} \cdot \frac{p_{i,j+1} -p_{i,j-1}}{2\Delta x}  \right. \nonumber \\
	\fl
			&& \left. -2q_{i,j}(1-q_{i,j}) \frac{ p_{i,j+1} - 2 p_{i,j} + p_{i,j-1}  }{(\Delta x)^2} \right] \label{eq:appBounceNum_disc_difeFirst} \\
	\fl
	\frac{p_{i+1,j } - p_{i-1,j}  }{2\Delta t}  &=& G(q_{i,j}, p_{i,j}) \nonumber \\
	\fl
			&-& \left[   \frac{ p_{i,j+1} - 2 p_{i,j} + p_{i,j-1}  }{(\Delta x)^2}  + (1-2q_{i,j})\frac{(p_{i,j+1}-p_{i,j-1})^2}{4(\Delta x)^2} \right]. 
\end{eqnarray}
For $i = 0, \ j \in \left\{ 0,1, \cdots, N_L -1\right\}$, 
\begin{eqnarray}
	\frac{p_{1,j } - p_{0,j}  }{\Delta t}  &=& G(q_{0,j}, p_{0,j}) \nonumber \\
			&-& \left[   \frac{ p_{0,j+1} - 2 p_{0,j} + p_{0,j-1}  }{(\Delta x)^2}  + (1-2q_{0,j})\frac{(p_{0,j+1}-p_{0,j-1})^2}{4(\Delta x)^2} \right].
\end{eqnarray}
For $i = N_T-1, \ j \in \left\{ 0,1, \cdots, N_L-1 \right\} $,
\begin{eqnarray}
	\frac{p_{N_T-1,j } - p_{N_T-2,j}  }{\Delta t}  &=& G(q_{N_T-1,j}, p_{N_T-1,j}) \nonumber \\
			&-& \left[   \frac{ p_{N_T-1,j+1} - 2 p_{N_T-1,j} + p_{N_T-1,j-1}  }{(\Delta x)^2}  \right. \nonumber \\
			&{}& \left. + (1-2q_{N_T-1,j})\frac{(p_{N_T-1,j+1}-p_{N_T-1,j-1})^2}{4(\Delta x)^2} \right].
			\label{eq:appBounceNum_disc_difeLast}
\end{eqnarray}
The original problem of solving partial differential equations is now reduced to the problem of finding zeros of 
a function $f: \mathbb{R}^{2N_L (N_T -1)} \rightarrow \mathbb{R}^{2N_L (N_T -1)}$. 
This problem can be solved numerically by Newton's method. 
Solving boundary value problems by applying Newton's method is called relaxation method \cite{num_recipes_C}.

\red{
Figure \ref{fig:bounce_q3} shows the calculated bounce solutions which determine the lifetime of $q_3$. 
\begin{figure}[tb]
	\centering
	\includegraphics[width=13cm]{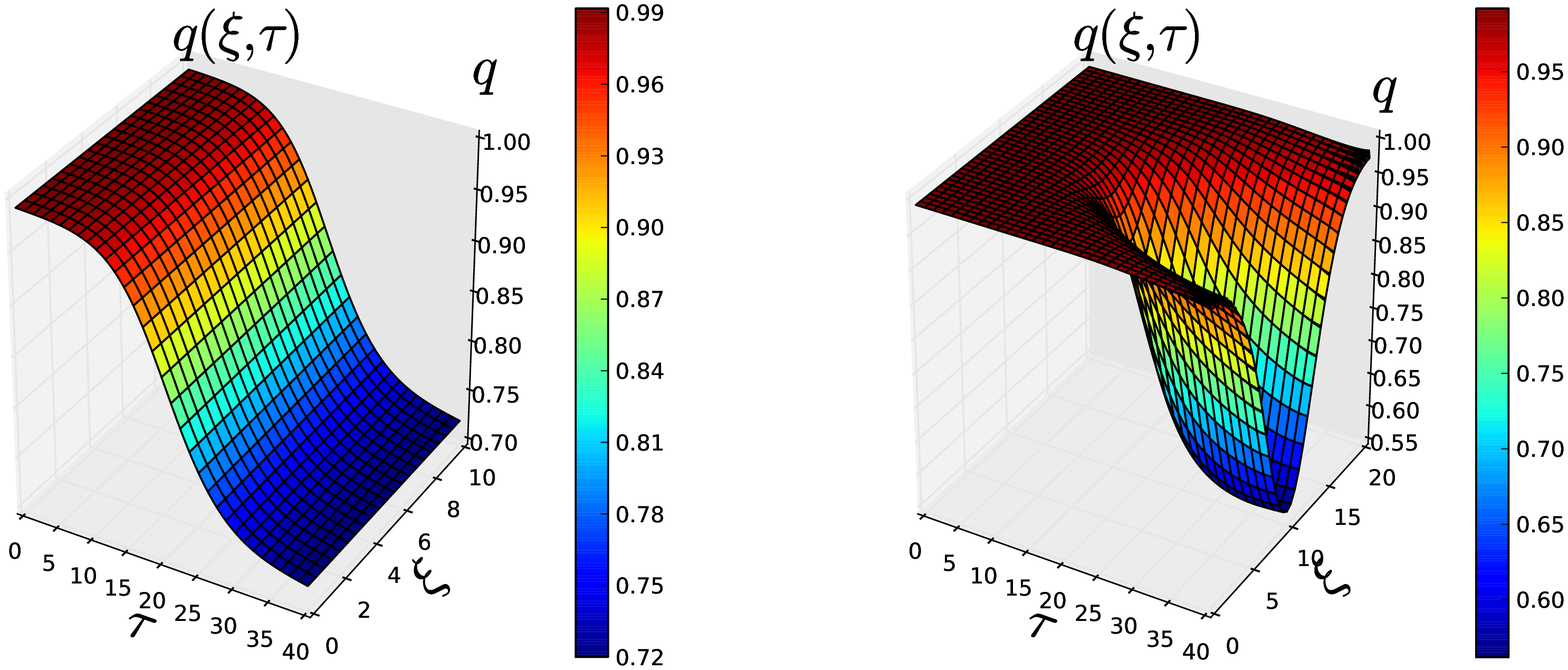}
	\caption{Bounce solutions which determine the lifetime of $q_3$. Left: uniform bounce solution $\alpha'$ ($L = 10.5$). Right: nonuniform bounce solution $\beta'$ ($L=20.5$).
		Parameters: $a-c=0.4, d-b=1.0, w = 0.8, \mu_A = \mu_B = 0.005$. 
		For this parameter set, $L_c \simeq 13.5$, and fixed points are $q_1 = 0.0031, q_2 = 0.72, q_3 = 0.99$.}
	\label{fig:bounce_q3}
\end{figure}
It can be easily seen from (\ref{eq:sp_stoch_SC_eqmq}) and (\ref{eq:sp_stoch_SC_eqmp}) that 
uniform bounce solution is the same as activation trajectory of the well-mixed model. 
This is because, if $q$ and $p$ do not depend on $\xi$, (\ref{eq:sp_stoch_SC_eqmq}) and (\ref{eq:sp_stoch_SC_eqmp}) coincide with 
the equations of motion for the well-mixed model (\ref{eq:wm_Stoch_SC_eqmq}) and (\ref{eq:wm_Stoch_SC_eqmp}), respectively. 
Bounce solutions which determine the lifetime of $q_1$ can be calculated in almost the same way 
by changing the boundary conditions, and are shown in figure \ref{fig:bounce_q1}.
\begin{figure}[tb]
	\centering
	\includegraphics[width=13cm]{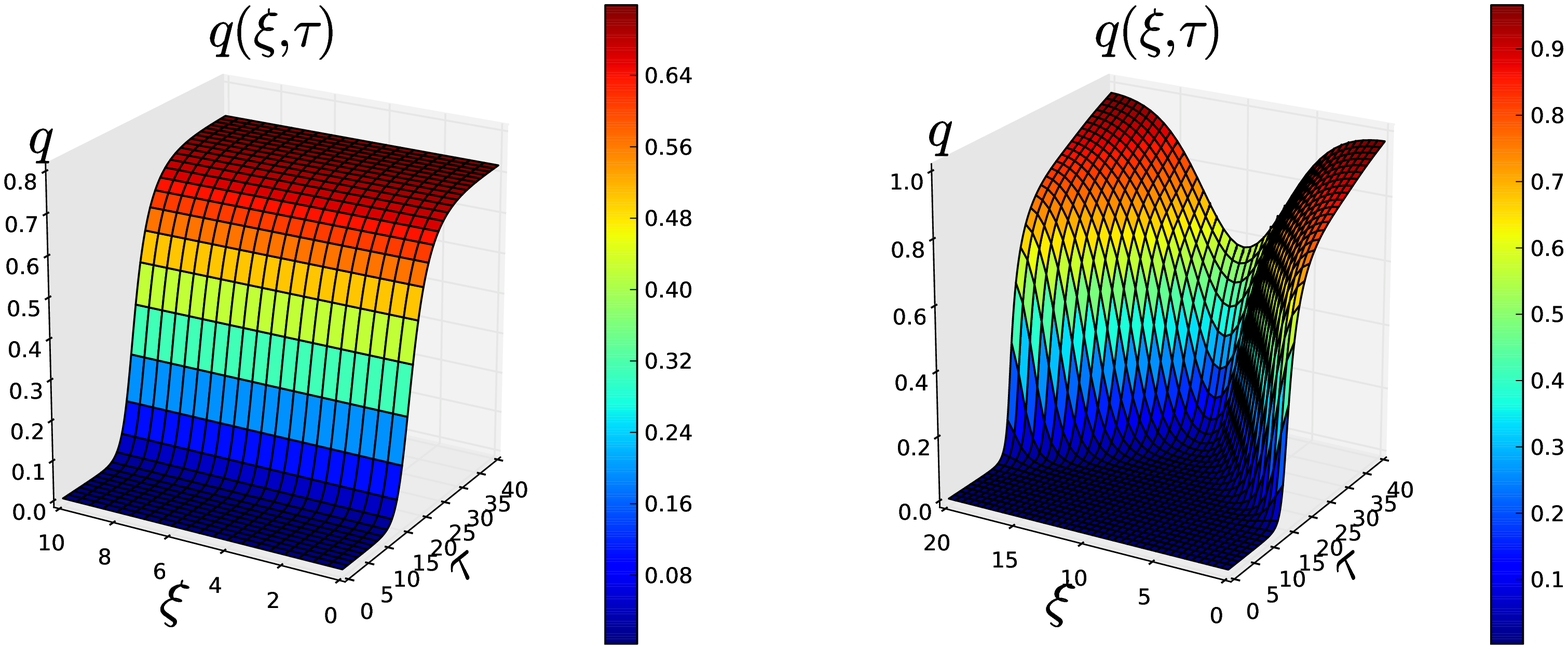}
	\caption{Bounce solutions which determine the lifetime of $q_1$. Left: uniform bounce solution ($L=10.5$). Right: nonuniform bounce solution ($L=20.5$).
		Parameters: $a-c=0.4, d-b=1.0, w = 0.8, \mu_A = \mu_B = 0.005$.
		For this parameter set, $L_c \simeq 13.5$, and fixed points are $q_1 = 0.0031, q_2 = 0.72, q_3 = 0.99$. }
	\label{fig:bounce_q1}
\end{figure}
}

\end{document}